\documentclass[aps,twocolumn,showpacs,preprintnumbers,amsmath,amssymb,nofootinbib,superscriptaddress,showkeys,prc]{revtex4-1}

\usepackage{hyperref}
\usepackage{float}
\usepackage{epsfig}
\usepackage{graphicx}
\usepackage{xcolor}
\usepackage{mathptmx}      
\usepackage{latexsym}
\usepackage{natbib}
\usepackage{longtable}
\usepackage{dcolumn}
\usepackage{comment}
\usepackage[utf8]{inputenc}

\DeclareMathVersion{nxbold}
\SetSymbolFont{operators}{nxbold}{OT1}{cmr} {b}{n}
\SetSymbolFont{letters}  {nxbold}{OML}{cmm} {b}{it}
\SetSymbolFont{symbols}  {nxbold}{OMS}{cmsy}{b}{n}

\DeclareMathAlphabet{\mathcal}{OMS}{cmsy}{m}{n}
\begin{document}

\title{Precise Determination of Charge Dependent Pion-Nucleon-Nucleon Coupling
  Constants}

\author{R. Navarro P\'erez}\email{navarroperez1@llnl.gov}
\affiliation{Nuclear and Chemical Science Division, Lawrence Livermore
  National Laboratory, Livermore, CA 94551, USA}

\author{J.E. Amaro}\email{amaro@ugr.es} \affiliation{Departamento de
  F\'{\i}sica At\'omica, Molecular y Nuclear \\ and Instituto Carlos I
  de F{\'\i}sica Te\'orica y Computacional \\ Universidad de Granada,
  E-18071 Granada, Spain.}
  
\author{E. Ruiz Arriola}\email{earriola@ugr.es}
\affiliation{Departamento de F\'{\i}sica At\'omica, Molecular y
  Nuclear \\ and Instituto Carlos I de F{\'\i}sica Te\'orica y
  Computacional \\ Universidad de Granada, E-18071 Granada, Spain.}

\date{\today}

\begin{abstract} 
We undertake a covariance error analysis of the pion-nucleon-nucleon
coupling constants from the Granada-2013 np and pp database comprising
a total of 6720 scattering data below LAB energy of 350 MeV. Assuming
a unique pion-nucleon coupling constant in the One Pion Exchange
potential above a boundary radius $r_c=3 {\rm fm}$ we obtain
$f^2=0.0763(1)$. The effects of charge symmetry breaking on the
$^3P_0$, $^3P_1$ and $^3P_2$ partial waves are analyzed and we find
$f_{p}^2 = 0.0761(4)$, $f_{0}^2 = 0.0790(9)$ and $f_{c}^2 = 0.0772(6)$
with a strong anti-correlation between $f_c^2$ and $f_0^2$. We
successfully test normality for the residuals of the fit. Potential
tails in terms of different boundary radii as well as chiral
Two-Pion-Exchange contributions as sources of systematic uncertainty
are also investigated.

\end{abstract}


\maketitle

\section{Introduction}

The meson exchange picture is a genuine quantum field theoretical
feature which implies, in particular, that the strong force between
protons and neutrons at long distances is dominated by the exchange of
the lightest hadrons compatible with the conservation laws, namely
neutral and charged pions. The strong force acting between nucleons at
sufficiently large distances or impact parameters $\gtrsim 3 {\rm
  fm}$ is solely due to one pion exchange (OPE) and was suggested by
Yukawa 80 years ago~\cite{Yukawa:1935xg}. The verification of this
mechanism not only provides a check of quantum field theory at the
hadronic level but also a quantitative insight onto the determination
of the forces which hold atomic nuclei~\cite{pauli1948meson}.  While
the mass of the pion may be determined directly from analysis of their
tracks or electroweak decays, the determination of the coupling
constant to nucleons needs further theoretical elaboration. The
pion-nucleon-nucleon coupling constant is rigorously defined as the
$\pi NN$ vertex function when all three particles are on the mass
shell and in principle any process involving the elementary vertices
$p \to \pi^0 p$, $n \to \pi^0 n$, $p \to \pi^+ n$ and $n \to \pi^- p$
(or their charge conjugated) is suitable for the determination of the
corresponding couplings provided all other relevant effects are
accounted for with an acceptable level of precision. In this work we
extract these coupling constants from NN scattering data and look for
signals of charge symmetry breaking.

The combinations entering in NN scattering are (we use the conventions
of \cite{Dumbrajs:1983jd} and when possible, for simplicity, omit the
$\pi$ label),
\begin{eqnarray}
  f_p^2 &=& f_{\pi^0 pp} f_{\pi^0 pp}  \\ 
  f_0^2 &=& -f_{\pi^0 nn} f_{\pi^0 pp} \\ 
  2 f_c^2 &=& f_{\pi^-pn} f_{\pi^+np} 
\end{eqnarray}
Usually the charge symmetry breaking is restricted to mass differences
by setting $f_{p} = -f_{n} = f_c = f_0 = f$.  The relevant
relationship between the pseudo-scalar pion coupling constant, $g_{\pi
  NN}$, and the pseudo-vector one, $f_{\pi NN}$, is given by
\begin{equation}
  \frac{g^2_{\pi^a N N'}}{4\pi} = \left( \frac{
    M_{N}+M_{N'}}{m_{\pi^\pm}} \right)^2 f^2_{\pi^a N N'}
\end{equation}
where $N,N'=n,p$ and $\pi^a= \pi^0, \pi^\pm$ (the factor $m_{\pi^\pm}$
is conventional). Thus, we may define $g_0^2$, $g_c^2$ and $g_p^2$. We
take $M_p=938.27231$ MeV the proton mass, $M_n=939.56563$ MeV the
neutron mass, and $m_{\pi^\pm}=139.5675$ MeV the mass of the charged
pion.

There is a long history of determinations of pion-nucleon coupling
constants using different approaches. A variety of methods and
reactions have been used since the seminal Yukawa paper. A more
complete account of the subsequent numerous determinations can be
traced from comprehensive overviews~\cite{deSwart:1997ep,
  Sainio:1999ba, Bugg:2004cm}. Here we will mainly review
determinations based on NN-scattering.

The very first determination was made in 1940 by Bethe by looking at
deuteron properties~\cite{PhysRev.57.260, Bethe:1940zz} soon after
Yukawa proposed his theory and before the pion was experimentally
discovered, finding the common value $f^2=0.077-0.080$. On a more
theoretical ground, based on dispersion relations and the Partial
Conservation of the Axial Current (PCAC) Goldberger and Treiman
deduced a relation between the $\pi NN$ form factor, $G_{\pi NN} (t)$,
the nucleon axial coupling constant, $g_A= 1.26$, and the pion weak
decay constant, $F_\pi=93.4(3)$ MeV. The relation, $G_{\pi NN} (0)
F_\pi = M_N g_A$ \cite{Goldberger:1958tr}, shown by Nambu to follow
from chiral symmetry~\cite{Nambu:1960xd}, is strictly valid at the
pion off-shell point, $q^2=0$, and numerically it yields $f_{\pi NN}^2
= g_A^2 m_{\pi^+}^2/(16\pi F_\pi^2) = 0.072$. Almost simultaneously,
Chew proposed \cite{Chew:1958zz} to determine it from the occurrence
of the pion pole in the renormalized Born approximation, by using an
extrapolation method which was implemented soon thereafter for
np~\cite{Cziffra:1959zza} and pp~\cite{MacGregor:1959zz} data. The
first direct and quantitative evidence for OPE was found in 1960 by
Signell~\cite{Signell:1960zz} by fitting the neutral pion mass to the
differential cross section in p-p scattering data. The method of
partial wave analysis (PWA) was soon afterwards used by Macgregor {\it
  et al.} \cite{macgregor1964determination}.

During many years $\pi N$ scattering determination through fixed-t
dispersion relations was advocated as a precision tool, yielding
initially $f_c^2 = 0.0790(10)$~\cite{Bugg:1973rv}, and later providing
$f_c^2=0.0735(15)$~\cite{Arndt:1990cn} (see also \cite{Arndt:1994bu}
and references therein). 
The latest most accurate $\pi N$ scattering
determinations are: 
i) the one based on the GMO rule
\cite{Ericson:2000md},
$g_c^2/(4\pi)=14.11(20)$ ($f_c^2=0.0783(11)$);
ii) the one 
using fixed-t
dispersion relations, $g_c^2/(4\pi)=13.76(8)$ \cite{Arndt:2006bf};
iii) the most recent one \cite{Baru:2010xn,Baru:2011bw}, based on
$\pi N$ scattering lengths, $\pi^- d$ scattering and the GMO sum
rule, yielding $g_c^2/(4\pi)=13.69(12)(15)=13.69(19)$.  Another source of
information has been the $\bar N N$ system, as shown by the Nijmegen
group \cite{Timmermans:1990tz}, providing $f_c^2=0.0751(17)$.

The modern era of high-quality NN interactions initiated by the
Nijmegen group~\cite{Stoks:1993tb} enabled to decrease the reduced
$\chi^2/\nu$ from 2 to 1, thanks to the implementation of charge
dependence (CD), vacuum polarization, relativistic corrections and
magnetic moments interactions, and a suitable selection criterion for
compatible data. Their analysis comprised a total of 4313 NN
scattering data.  This promoted the determination of the pion-nucleon
coupling constant from np and pp scattering to a competitively
accurate approach. The main advantage of an NN analysis as compared to
the $\pi N$ analysis, which has so far been restricted to charged
pions, is that one can determine both neutral and charged-pion
coupling constants simultaneously, to search for isospin breaking
effects. The three compatible values, $f_p^2=0.0751(6)$,
$f_0^2=0.0752(8)$ and $f_c^2=0.0741(5)$, were determined from NN
scattering data~\cite{Klomp:1991vz}. The originally recommended charge
independent value $f^2=0.0749(4)$~\cite{Klomp:1991vz} was
revised~\cite{Stoks:1992ja} and confirmed in the 1997 review on the
status of the pion-nucleon-nucleon coupling
constant~\cite{deSwart:1997ep}; this is the most accurate NN
determination to date.  There, it was suggested that a
charge-independence breaking could be checked with more data and
better statistics. The most recent determinations of the Nijmegen
group have been given after the inclusion of charge-independent chiral
two-pion exchange ($\chi$TPE) potential~\cite{Kaiser:1997mw} which
depends on three additional chiral constants, $c_1$, $c_3$, $c_4$,
which also appear in $\pi N$ scattering. A combined fit of $f_p^2$ and
$c_{1,3,4}$ to pp scattering data, provides the value
$f_p^2=0.0756(4)$~\cite{Rentmeester:1999vw}, and a simultaneous fit to
pp+np data of a common $f^2$ and $c_{1,3,4}$ \cite{Rentmeester:2003mf}
provides linear correlations between $f^2$ and $c_{1,2,3}$.

Most of the analyses determining the pion nucleon coupling constants
involve heavy statistical analysis for a large body of experimental
data, mostly $\chi^2-$fits, which are subjected to a number of {\it a
  posteriori} tests~\cite{evans2004probability}. The verification of
these tests buttress a sensible analysis of uncertainties of
theoretical models~\cite{Dobaczewski:2014jga}.  To the time of their
analysis, the Nijmegen group~\cite{Bergervoet:1988zz,Stoks:1992ja}
checked the statistical quality of pp fit residuals using the moments
test, which for increasing orders over-weights the tails.

In this paper we study the possible differences among the pion-nucleon
coupling constants by analyzing np and pp scattering data using the NN
Granada-2013, $3\sigma$-self consistent database, designed and
analyzed recently~\cite{Perez:2013cza, Perez:2013jpa, Perez:2014yla,
  Perez:2014kpa}~\footnote{The 2013 Granada database is available at
  \url{http://www.ugr.es/~amaro/nndatabase/}}.  There, we have
selected 6713 out of 8000 published np and pp experimental data for
LAB energies below 350 MeV and measured in the period 1950-2013, which
satisfactorily verify the tail-sensitive test~\cite{Aldor2013}, based
on the quantile-quantile plot for the combined np+pp residuals (see
also Ref.~\cite{Perez:2015pea} for an application of these ideas to
$\pi\pi$ scattering). As a side remark we note that the Uppsala
controversial measurement \cite{Ericson:1995gr,Rahm:1998jt}, which
gives the value $f^2=0.081$ and appears in Weinberg's textbook
\cite{Weinberg:1996kr}, was disputed by the Nijmegen
group~\cite{Rentmeester:1998vf} and
contested~\cite{Ericson:1998fq}. An overview of the situation can be
glanced in~\cite{Blomgren:2000wq}. This measurement has been rejected
by our $3\sigma$ self-consistent database~\cite{Perez:2013jpa}. A
latter re-measurement at IUCF by the partly the same
group~\cite{Sarsour:2004xx,Sarsour:2006fd}, which is compatible with
the original Nijmegen PWA, is not rejected by the self-consistent
$3\sigma$ criterion.

The paper is organized as follows. In Section~\ref{sec:CD-OPE} we
describe the OPE potential, introduce our notation, and discuss the
conditions under which we naturally expect to unveil charge dependence
in the pion-nucleon coupling constants. In Section~\ref{sec:Granada}
we review the main aspects of our partial wave analysis and the
Granada-2013 database. Our motivation for incorporating charge
dependence in the P-waves, besides the customary charge dependence on
S-waves implemented in all modern high quality fits, is presented in
Section~\ref{sec:numerics} along with a discussion of our numerical
results based on a covariance analysis. An effort to quantify
systematic errors by analyzing the long range component of the CD-OPE
is made in Section~\ref{sec:system}. Finally, in
Section~\ref{sec:conclusions} conclusions are presented. In the
Appendix we show the extended operator basis accommodating S-wave and
P-wave charge dependence.

\section{Charge-Dependent One Pion Exchange}
\label{sec:CD-OPE}

The charge-dependent, one-pion exchange (CD-OPE) potential incorporates
charge symmetry breaking by considering the mass difference of the
neutral and charged pions as well as assuming different coupling
constants. We use the convention for $\pi NN$ Lagrangians defined in
the review of ref. \cite{Dumbrajs:1983jd}. The quantum mechanical potential
which reproduces in Born approximation 
the corresponding Feynman diagrams for on-shell
static nucleons, is given in the pp, nn and np
channels as
\begin{eqnarray}
 V_{{\rm OPE},pp}(r) &=& f^2_{p} V_{m_{\pi^0},\rm OPE}(r),  \\
 V_{{\rm OPE},nn}(r) &=& f^2_{n} V_{m_{\pi^0},\rm OPE}(r),  \\
 V_{{\rm OPE},np}(r) &=& -f_{n}f_{p}V_{m_{\pi^0},\rm OPE}(r)- (-)^{T} 2f^2_cV_{m_{\pi^\pm},\rm OPE}(r), 
 \label{eq:BreakIsospinOPE}
\end{eqnarray}
respectively. Here, $V_{m,\rm OPE}$ is given by
\begin{equation}
 V_{m, \rm OPE}(r) =
 \left(\frac{m}{m_{\pi^\pm}}\right)^2\frac{1}{3}m\left[Y_m(r){\mathbf
     \sigma}_1\cdot\mathbf{\sigma}_2 + T_m(r)S_{1,2} \right].
\end{equation}
Here $Y_m$ and $T_m$ are the usual Yukawa functions,
\begin{eqnarray}
 Y(r) &=& \frac{e^{-m r}}{m r} \\  
 T(r) &=& \frac{e^{-m r}}{m r } \left[ 1 + \frac3{mr} + \frac{3}{(mr)^2}\right],
\end{eqnarray}
$\sigma_1$ and $\sigma_2$ are the single nucleon Pauli matrices, and
$S_{12}= 3 \sigma_1 \cdot {\bf \hat r} \sigma_2 \cdot {\bf \hat r}
-\sigma_1 \cdot \sigma_2 $ is the tensor operator.  Unfortunately, the
CD-OPE potential by itself cannot be directly compared to experimental
data, and the only way we know how to determine these pion-nucleon
couplings is by carrying out a PWA.

From a purely classical viewpoint, in order to {\it measure} the
nuclear force directly it would just be enough to hold and pull two
nucleons apart at distances larger than their elementary size, which
is or the order of $2 {\rm fm}$~\cite{Perez:2013cza}. For such an
ideal experiment the behavior of the system at shorter distances would
be largely irrelevant, because nucleons would behave as point-like
particles. This situation would naturally occur if nucleons were
truly infinitely heavy. In that case the potential would correspond to
the static energy of a system with baryon number $B=2$ and total
charge $Q=2,1,0$, for $pp$, $pn$ 
$nn$, 
respectively\footnote{This is the case in lattice calculations, where static
  sources are placed at a fixed
  distance~\cite{Aoki:2011ep,Aoki:2013tba}.  In the quenched
  approximation it has been found, for a pion mass of $m_\pi=380$
  MeV, the value $g^2/(4\pi)=12.1\pm 2.7$, which is
  encouraging~\cite{Aoki:2009ji} but still a crude estimate.}. Of
course, the quantum mechanical nature of the nucleons prevents such a
situation experimentally and we are left with scattering experiments.
Good operating conditions are achieved when the maximum relative CM
momentum, $p_{\rm max}$, is small enough to avoid complications due to
inelastic channels and large enough to contain as many data as
possible.  This generates a resolution ambiguity of the order of the
minimal relative de Broglie wavelength, $\lambda_{\rm min} =\Delta r
\sim 1/p_{\rm max}$. Since the $NN \to \pi NN$ channel opens up at
$p_{\rm max}\sim \sqrt{m_\pi M_N}\sim 360 {\rm MeV}$, we have $\Delta
r \sim 0.6 {\rm fm}$. Unfortunately, in the quantum mechanical NN
scattering problem the scales are somewhat intertwined, and thus some
information on the unknown short-distance components of the
potential have to be considered in order to evaluate
 the scattering amplitude, the cross
section or the polarization asymmetry. 
The low energy behavior of the NN interaction is expected to 
depend strongly on its long
distance properties. Although some coarse grained
information of the unknown contribution is needed, it can
be deduced from the experiment with an overall {\it sufficient}
accuracy as to determine the differences between the pion-nucleon couplings.
This viewpoint allows to determine {\it a priori} the number of
independent parameters $N_{\rm Par}$ needed for a successful
fit\footnote{In \cite{Perez:2013cza} it was found that, for $r_c=3
  {\rm fm}$ , the number of needed parameters is 
$N_{\rm Par} \sim 60$. The argument is based on the idea
  that, if we adopt the CD-OPE potential above $r_c$, we can estimate the
  number of independent potential values $V(r_n)$ below $r_c$ in any
  partial wave channel, with $r_n = n \Delta r$. Since the maximum
  angular momentum in the partial wave expansion is $l_{\rm max} \sim
  p_{\rm max} r_c$ and there are four independent waves for each $l$, we
  would have $N_{\rm Par} \sim 4 l_{\rm max} (r_c /\Delta
  r)$. Excluding the points $r_n$ below the centrifugal barrier, the
  number becomes $N_{\rm Par} \sim 2 (p_{\rm max}r_c )^2$.}. These
ideas where introduced by Aviles long ago~\cite{Aviles:1973ee}, and
they underlie the recent NN analysis carried out by the present authors, 
where a large database, comprising about 8000
published experimental data measured in the period 1950-2013, was
considered~\cite{Perez:2013cza,Perez:2013jpa}.

\subsection{The number of data}

There is no symmetry reason why the strong force between protons and
between neutrons should be exactly identical; if a difference exists
one should be able to see it with a sufficiently large amount of
experimental data.  These differences are in fact small and hard to
pin down because {\it a priori} the electromagnetic corrections should scale
with the fine structure constant $ \delta g/g \sim \alpha \sim 1/137$,
and the strong (QCD) corrections should scale with the $u-d$ quark mass
difference (relative to the $s$-quark mass) which means $\delta g /
g\sim( m_u-m_d )/\Lambda_{\rm QCD} \sim (M_p-M_n)/\Lambda_{\rm QCD}
\sim 1/100 $, for $\Lambda_{\rm QCD} \sim 250 {\rm MeV}$. This simple
estimates suggest that in order to witness isospin violations in the
couplings we should determine them with a target accuracy better than
$1-2\% $, which is not too far from the most recent values. On a
purely statistical basis the relative uncertainty due to $N$
independent measurements is $ \Delta g/g \sim 1/\sqrt{N}$. If we have
some extra parameters $(\lambda_1, \dots, \lambda_{N_{\rm Par}})$,the
condition $\Delta g \sim \delta g \sim 0.01-0.02$ would require
$N=N_{\rm Dat}-N_{\rm Par} \sim 7000-10000$ {\it independent} degrees
of freedom. Since $N_{\rm Dat} \gg N_{\rm Par} $ this is comparable to
the total amount of existing elastic np and pp scattering data. While
these are rough estimates, we stress the independence character of the
measurements in order to make these estimates credible; it is not just
a question of having more data. From the point of view of
$\chi^2$-fits this requires passing satisfactorily normality tests
guaranteeing the self-consistency of the fit.  In particular, adding
many incompatible data would invalidate this analysis.

\subsection{Naturalness of fitting parameters}

While our approach is based on a standard least squares optimization,
which minimizes the distance between the theory and the experiment for
many pp and np scattering data, it is important to mention that we do
not consider that all fits are eligible and in fact some of them will
be rejected. In what follows, we specify these criteria {\it a
  priori}.

As a matter of principle, we reject fits which display bound states in
channels other than the deuteron (occurring in the $^3S_1-^3D_1$
channel only) which will be considered spurious. The appearance of
such states in the fitting process is not so unlikely, particularly in
the case of peripheral waves. This is usually detected by use of
Levinson's theorem, $\delta_l(0)-\delta_l(\infty)= n \pi$, which
requires checking phases at energies much larger than the fitting
range. An equivalent way to find spurious bound states is by
checking the volume integrals (and high moments) for which a large degree
of universality has been found~\cite{Perez:2016vzj}.  Too large
attractive couplings in the potential permitting a bound state would
consequently generate unnaturally large volume integrals of the
potentials.

In the case of the pion-nucleon coupling constant we expect some
theoretical constraints to be fulfilled. The renowned
Goldberger-Treiman relation was deduced as a consequence of exact
PCAC, and yields the value (in the isospin limit)
\begin{eqnarray}
  G_{\pi NN} (0) = M_N g_A /F_{\pi} 
\end{eqnarray}
The physical coupling constant corresponds to $g_{\pi NN} \equiv
G_{\pi NN} ( m_\pi^2) > G_{\pi NN} (0)$. It is expected to be {\it
  larger} than value appearing in GT-relation. This suggests
\begin{eqnarray}
  g_{\pi NN} > M_N g_A /F_{\pi}
\end{eqnarray}
and hence, for the PDG values $F_{\pi^+}^{\rm PDG}= 92.21(14) {\rm
  MeV}$ and $g_A=1.2723(23)$,
\begin{eqnarray}
  f_{\pi NN}^2 > f_{\pi NN, {\rm GT}}^2 \equiv \frac1{16 \pi}
  \left(\frac{g_A m_{\pi^+}}{F_\pi}\right)^2 =0.07324(4)
  \label{eq:golberger-treiman}
\end{eqnarray}
The uncertainty is about $5\%$.  More generally, a GT-discrepancy is
defined (see e.g. \cite{Dominguez:1984ka} for a review),
\begin{eqnarray}
  \Delta_{\rm GT}= 1- \frac{M_N g_A}{g_{\pi NN} F_\pi}
\end{eqnarray}
The value of this number has been changing but typical values nowadays
are at the few percent level, $\Delta_{\rm GT} \sim 0.01-0.03$.  In
the limit of zero quark masses, chiral symmetry becomes exact, and
hence $\Delta_{\rm GT} = {\cal O} (m_\pi^2/F_\pi^2)$

The fact that $(m_u-m_d)/(m_u+m_d) \sim 1/3$ suggests that, if we
obtain a GT discrepancy different from zero, about three more times
precision would be needed to pin down isospin breaking. According to
our $1/\sqrt{N_{\rm Dat}}$ estimate above, this can be accomplished by
increasing the number of independent data by a factor of 10.  At the
level of isospin breaking some estimates have also been
made~\cite{Goity:1999by,Goity:2002uh}.
 
In the case of $\chi$TPE exchange, which will also be considered
below, the chiral constants $c_{1,3,4}$ are saturated by meson
exchange~\cite{Bernard:1996gq}. Actually, $c_1$ is saturated by scalar
exchange.  The saturation value is $c_1^S = - g_S c_m /m_S^2 $. Taking
$M_N = g_S F_\pi $ and $c_m = F_\pi/2$ , $m_S=m_V = F_\pi \sqrt{24 \pi
  /N_c}$~\cite{Ledwig:2014cla} and $M_N = N_c m_\rho/2$ we get $c_1^S
\sim -N_c /( 4 \sqrt{2} m_\rho) \sim -0.7 {\rm GeV}^{-1}$. In the case
of the constants $c_3$ and $c_4$, they are saturated by $\Delta$
resonance; taking $\Delta = M_\Delta -M_N$, the saturation values are
$c_2^\Delta = -c_3^\Delta = 2 c_4^\Delta = g_A^2/(2 \Delta) \sim 2.97
{\rm GeV}^{-1}$. Of course, these are not very accurate values, but
indicate the order of magnitude one should expect.

\section{The Granada-2013 analysis}
\label{sec:Granada}

In a series of works we have upgraded the NN database to include a
total of 6720 np and pp published experimental data by using a coarse
grained representation of the interaction, and applying stringent
statistical tests on the residuals of the $\chi^2$-fits after 
implementing
a $3\sigma$ self consistent selection process
\cite{Perez:2014yla}. The resulting Granada-2013 is at
present the largest NN database which can be described by a CD-OPE
contribution. There are about $60\%$ more data than the 4313 data used in
the latest Nijmegen upgrade~~\cite{deSwart:1997ep}. This suggests that we
can improve on the errors for the pion-nucleon couplings as discussed
in the previous section.

We have discussed in detail the many issues in carrying out the data
selection, the fit and the corresponding joint np+pp partial wave
analysis. We review here the main aspects as a guideline and refer to
those works for further details.

We separate the potential into two well defined regions depending on a
chosen cut-off radius, $r_c$, fixed in such a way that for $r> r_c$ the
CD-OPE is the only strong contribution. In addition, for $r>r_c$ we
also include electromagnetic (Coulomb, vacuum polarization, magnetic
moments) \cite{Perez:2013jpa} and relativistic corrections which we simply
add to the strong potential.
\begin{equation}
  V({\bf r})= V_{\rm OPE}({\bf r}) + V_{\rm EM}({\bf r}),
  \kern 1cm r > r_c
\end{equation}
Below the cut-off radius, $r< r_c$ we regard the NN force as unknown,
and we use delta-shells located at equidistant points separated by
$\Delta r =0.6 {\rm fm}$, corresponding to the shortest de Broglie
wavelength at  pion production threshold.  The fitting parameters
are the real coefficients $(\lambda_i)^{JS}_{ll'}$ for each partial
wave:
\begin{equation}
  V_{l,l'}^{JS}(r) = \frac{1}{2\mu}\sum_{i=1}^N (\lambda_i)^{JS}_{ll'}
  \delta(r-r_i), \kern 1cm r \leq r_c.
\end{equation}
where $\mu$ is the NN reduced mass.  Alternatively the potential can
be expanded in an operator basis extending the AV18 potentials in
coordinate space, see appendix A.  The transformation between
partial wave and operator basis was given in
Ref.~\cite{Perez:2013jpa}.

It turns out that $r_c=3 {\rm fm}$ provides statistically satisfactory
fits to the selected $3\sigma$-self consistent Granada-2013 database.
While it would be interesting to separate explicitly the known from
the unknown pieces of the interaction below the cut-off radius $r_c$,
this is actually a complication in the fitting procedure, and will not
change the values of the most-likely pion-nucleon coupling constants.
Another advantage of taking $r_c=3 {\rm fm}$ is that in our analysis
there is no need of form factors of any kind, and thus we are relieved
from disentangling finite size effects, quark exchange and the
intrinsic resolution $\Delta r$ inherent to any finite energy
PWA~\footnote{An Explanation of the Apparent Charge Dependence of the
  Pion Nucleon Coupling was attributed to the strong form
  factor~\cite{Thomas:1989tv}.}

The possible $A_y$ problem for np scattering, raised by the data of
Ref.~\cite{Braun:2008eh}, suggested a sizable isospin breaking of
coupling constants. The problem was re-analyzed theoretically in ref. 
\cite{Gross:2008pd}, and motivated the reanalysis of the
data~\cite{Weisel:2010zz} and the disentanglement between
systematic and statistical errors.  Actually, in ref. \cite{Gross:2008pd}
it was found that these data might be explained in an isolated fashion when
isospin was broken. Thus, we allow this isospin breaking to foresee the
possibility of recovering the data.

\begin{figure}[tb]
\begin{center}
\epsfig{figure=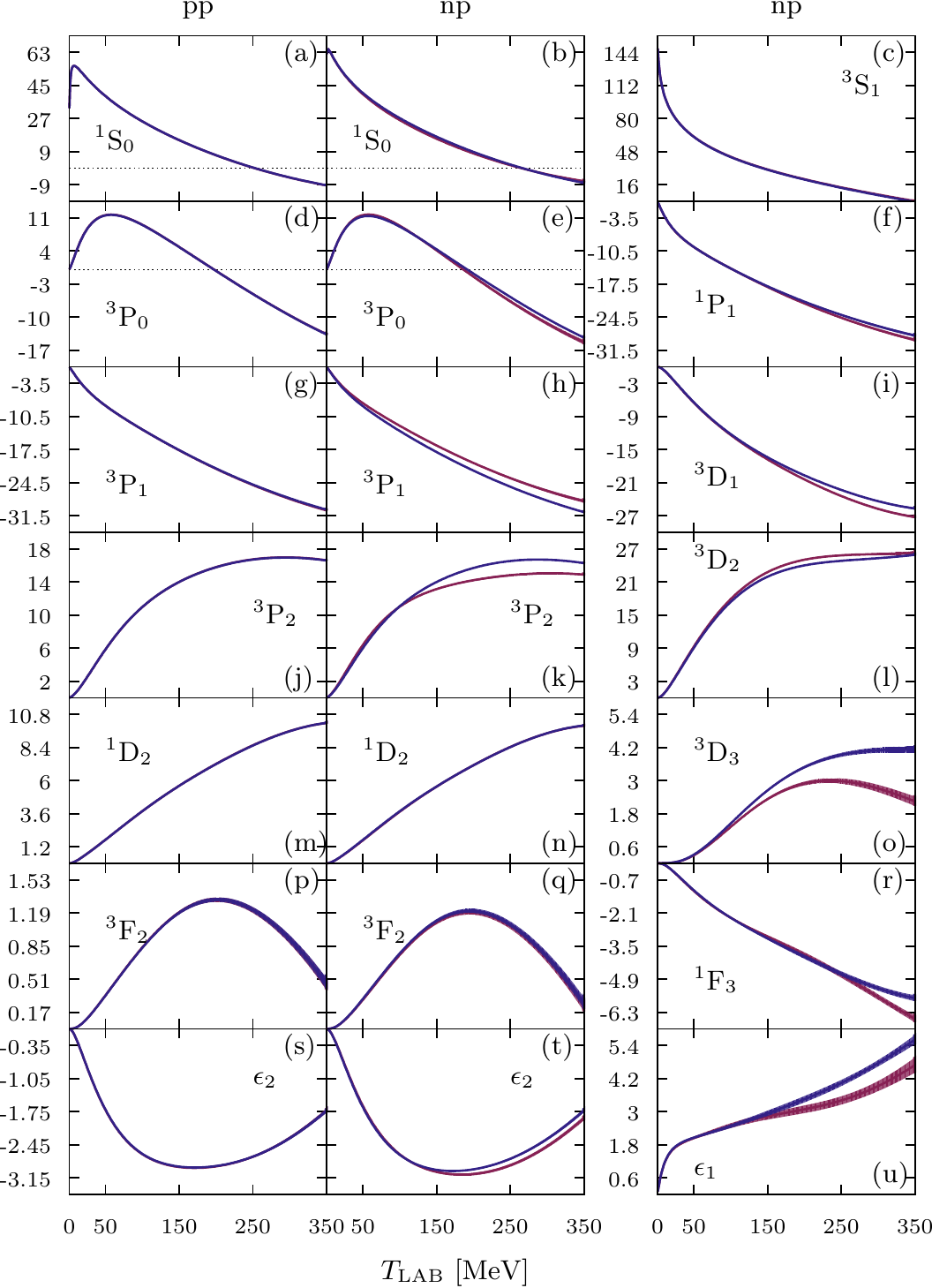,width=1\linewidth} \end{center}
\caption{(Color online) Phaseshifts obtained from a partial waves
  analysis to pp and np data and statistical uncertainties. Blue band
  from~\cite{Perez:2013jpa} and red band from a fit with charge
  symmetry breaking on the $^3P_0$, $^3P_1$ and $^3P_2$ partial
  waves.}
\label{Fig:Phaseshifts}
\end{figure}

\section{Statistical analysis}
\label{sec:numerics}

In our previous analysis we took a fixed common value for the
pion-nucleon coupling constant suggested by the Nijmegen group.  When
we relax this assumption and also fit the pion-nucleon coupling
  constant as another parameter in the potential, we obtain
$f^2=0.0763(1)$, which is $3\sigma$ compatible with the Nijmegen
recommendation \cite{Klomp:1991vz}, $f^2=0.0749(4)$ and more accurate.

\subsection{Charge symmetry breaking on $S$- and $P$ waves}

An old problem in NN scattering fitting is if it is possible to
predict the neutron-neutron potential fron np and pp data.  A
necessary condition would be that the unknown piece of the short
distance interaction for np and pp coincide in the isovector
channels. Once we allow to vary the coupling constants $f_p^2$,
$f_0^2$ and $f_c^2$ from their common value $f^2$ we have first
searched for a fit {\it without} CD in the $\lambda's$ (i.e. assuming
that they are equal for np and pp).  We get $\chi^2/\nu = 1.2 $ for
CD-OPE above $r_c=3 {\rm fm}$. On the other hand, $\chi^2/\nu = 9 $
for CD-OPE+$\chi$TPE above $r_c=1.8 {\rm fm}$. Therefore, and in
harmony with all high-quality previous attempts we cannot deduce
nn-scattering below $r_c= 3{\rm fm}$.

Following the common practice of other
analyses~\cite{Stoks:1993tb,Wiringa:1994wb,Machleidt:2000ge} , we have
previously allowed different pp and np parameters only on the $^1S_0$
partial
wave~\cite{Perez:2013cza,Perez:2013oba,Perez:2013jpa,Perez:2013mwa}
and found that this symmetry breaking is indeed necessary to obtain an
accurate description of the pp and np scattering data. The large
collection of about $8000$ available data also makes it possible to
test charge symmetry breaking on the parameterization of higher
partial waves, e.g. $^3P_0$, $^3P_1$ and $^3P_2$.

To carry out such a test we have considered different np and pp
parameters on those partial waves and performed a full PWA and
selection process as described in~\cite{Perez:2013mwa,Perez:2013jpa},
by fitting the delta-shell potential parameters to the complete
database and then applying the $3\sigma$ rejection criterion
iteratively until a self consistent database is obtained. The
consistent database obtained in this case has $3006$ pp data and
$3735$ np data, including normalizations, and the value for the chi
square per number of data is $\chi^2/N_{\rm data} = 1.02$. When
comparing with our previous consistent data base~\cite{Perez:2013jpa}
this symmetry breaking can only describe $21$ additional data out of
more than $1000$ rejected data. Fig.~\ref{Fig:Phaseshifts} compares
the low angular momentum phaseshifts of the PWA
in~\cite{Perez:2013jpa} (blue bands) with this new analysis (red
bands). The pp phaseshifts show no significant difference, while the
np ones are statistically different and the differences are even
greater for higher angular momentum partial waves. Tabulated values
for the lower phase-shifts for selected LAB energies are provided in
Appendix~\ref{sec:phases}.

\begin{table*}[tb]
  \caption{\label{tab:fpf0fc} The pion-nucleon coupling constants $f_p^
    2$, $f_0^ 2$ and $f_c^ 2$ determined from different fits to the
    Granada-2013 database and their characteristics. We indicate the
    partial waves where charge dependence is allowed.}
  \begin{tabular*}{\textwidth}{@{\extracolsep{\fill}}cccccccccc}
    \toprule
    $f_p^2$ & $f_0^2$ & $f_c^2$ & CD-waves & $\chi^2_{pp}$ &  $\chi^2_{np}$ & $\chi^2$ & $N_{\rm Dat}$ & $N_{\rm Par}$ & $\chi^2/\nu $ \\ 
    \colrule
    0.075     & idem      & idem      & $^1S_0$      & 2997.29 & 3957.57 & 6954.86 & 6720 & 46 & 1.042 \\ 
    0.0763(1) & idem      & idem      & $^1S_0$      & 2995.20 & 3952.85 & 6947.05 & 6720 & 47 & 1.041 \\ 
    0.0764(4) & 0.0779(8) & 0.0758(4) & $^1S_0$      & 2994.41 & 3950.42 & 6944.83 & 6720 & 49 & 1.041 \\
    0.0761(4) & 0.0790(9) & 0.0772(5) & $^1S_0$, $P$ & 2979.37 & 3876.13 & 6855.50 & 6741 & 55 & 1.025 \\  
    \botrule
  \end{tabular*}
\end{table*}

Usually the charge symmetry breaking is restricted to mass differences
by setting $f_{p} = -f_{n} = f_c = f$. The value $f^2 = 0.075$
recommended by the Nijmegen group~\cite{Klomp:1991vz} has been
used in most of the potentials since the seminal 1993 partial wave
analysis~\cite{Stoks:1993tb}.  Here we test this charge independence with
the large body of data available today, by using $f_{p}$, $f_{0}$, and
$f_{c}$ as extra fitting parameters along with the previous $46$
delta-shell parameters.
We show our results in Table~\ref{tab:fpf0fc} depending on different
strategies regarding isospin breaking:  S-waves,  S- and P-waves, 
and in the coupling constants. The working group summary of 1999
provides a recent compilation of coupling constants in a chronological
display~\cite{Sainio:1999ba}. The most recent
determination~\cite{Baru:2010xn,Baru:2011bw}, based on $\pi N$
scattering lengths and $\pi^- d$ scattering, and in the GMO sum rule,
yields $g_c^2/(4\pi)=13.69(12)(15)=13.69(20)$. From our full
covariance matrix analysis we get $g_p^2/(4\pi)= 13.774(75)$,
$g_0^2/(4\pi)=14.30(16)$ and $g_c^2/(4\pi)=13.984(82)$. The last value
is $2\sigma$ compatible with these determinations, but slightly more
accurate.

\begin{figure*}[htb]
  \centering
  \includegraphics[width=0.33\textwidth]{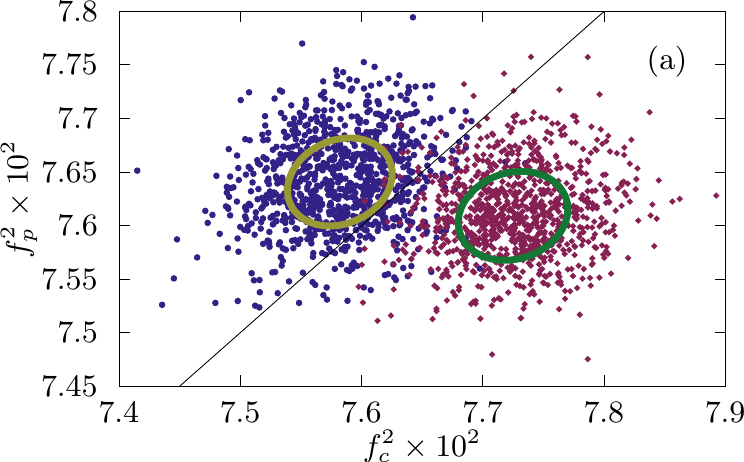}
  \includegraphics[width=0.33\textwidth]{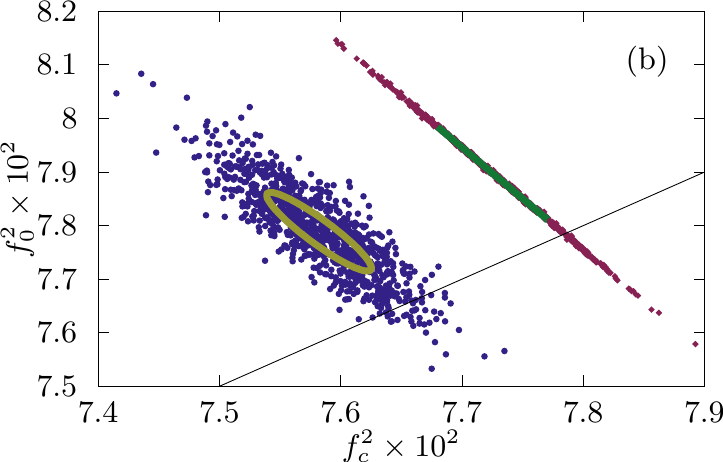}
  \includegraphics[width=0.33\textwidth]{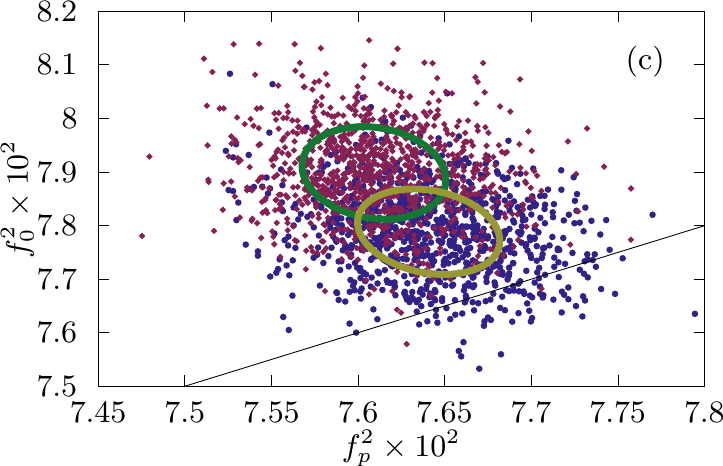}
  \caption{(Color Online) Correlation ellipses and scatter diagrams
    for the coupling constants $f_c^2$ , $f_{p}^2$ and $f_{0}^2$
    appearing in the OPE potential from a PWA with (yellow line and
    blue dots) and without (green line and red diamonds) charge
    independence on the $P$ waves and a $3\sigma$ consistent
    database. The black diagonal line indicates $f_c^2=f_p^2=f_0^2$}
  \label{fig:fcp0ellipses}       
\end{figure*}

\begin{figure*}[htb]
  \centering
  \includegraphics[width=\textwidth]{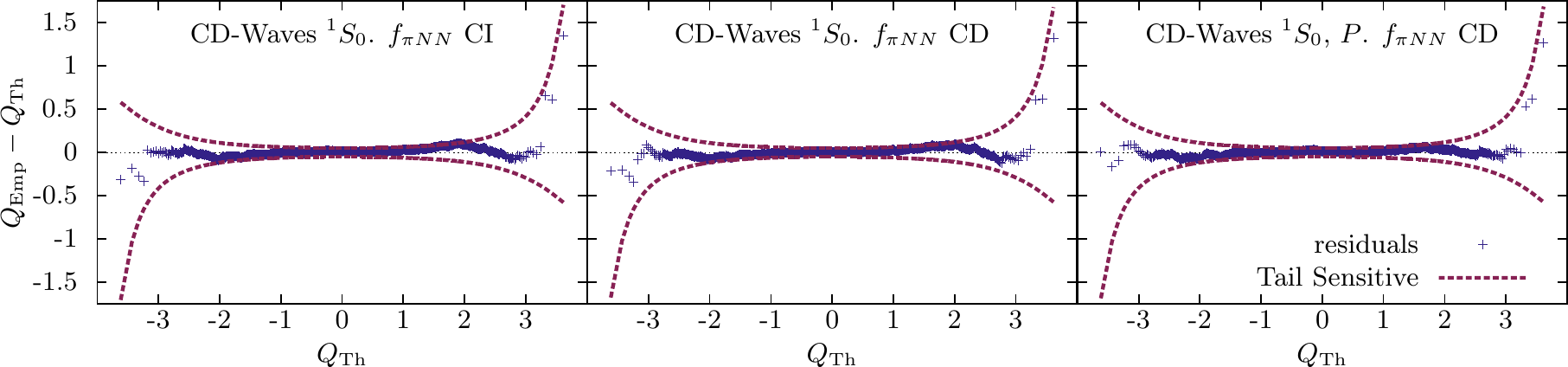}
  \caption{(Color Online) Rotated quantile-quantile plots for the fits
    introduced in this work. All points should be inside the
    confidence band to state that residuals of the fit follow a normal
    distribution $N(0,1)$, in which case the fit is self-consistent
    {\it a posteriori}. Left panel, assuming a charge independent
    pion-nucleon constant used as a fitting parameter and charge
    symmetry breaking only on the $^1S_0$ partial wave
    parameters. Central panel, assuming three different charge
    dependent pion-nucleon constants used as a fitting parameters and
    charge symmetry breaking only on the $^1S_0$ partial wave
    parameters. Right panel, assuming three different charge dependent
    pion-nucleon constants used as a fitting parameters and charge
    symmetry breaking on the $^1S_0$ and $P$ partial wave parameters }
  \label{fig:qqplots}       
\end{figure*}

The fitting delta-shell parameters obtained in our different
strategies, regarding charge independence breaking in just S-waves and
charge independence breaking in S and P waves, can be seen in
Tables~\ref{tab:FitParameters} and \ref{tab:FitParametersPwaves}
respectively. We use the resulting parameters along with their
covariance matrix to calculate $f_{p}^2$, $f_0^2 $ and $f^2_c$, and
propagate the corresponding statistical uncertainties and test charge
dependence. Fig.~\ref{fig:fcp0ellipses} shows the $1\sigma$
correlation ellipses along with the scatter diagram resulting from
drawing $1000$ random variates following the multivariate normal
distribution dictated by the covariance matrix. The fit without charge
dependence on the $P$ waves is indicated by the blue dots and yellow
line while the fit with charge dependence on the $P$ waves corresponds
to the red diamonds and green line. Charge independence, $f_p^2 = f_0^2
= f_c^2$, is marked by the diagonal black line. Several aspects should
be noted from Fig.~\ref{fig:fcp0ellipses}. First, while the values on
Table~\ref{tab:fpf0fc} seem to suggest that the determinations with
and without charge charge dependent $P$ waves for $f_0^2$ and $f_c^2$
are $1$ and $2\sigma$ compatible respectively, in fact the strong
anti-correlation between the two coupling constants makes the
determinations completely incompatible. The determination with charge
dependence on the $S$ waves only is compatible with the  $f_p^2 = f_0^2
= f_c^2 = 0.0763(1)$ fit at the two sigma level; this is in accordance
with the slight decrease in $\chi^2$ in spite of the fact that two
extra parameters are fitted. Finally, the fit with charge dependent
$P$ waves is incompatible with $f_p^2 = f_0^2 = f_c^2$, once again due
to the strong anti-correlation between $f_0^2$ and $f_c^2$.

\begin{table}[tb]
  \caption{\label{tab:FitParameters} Fitting delta-shell parameters
    $(\lambda_n)^{JS}_{l,l'} $ (in ${\rm fm}^{-1}$) with their errors
    for all states in the $JS$ channel for a fit with isospin symmetry
    breaking on the $^1S_0$ partial wave parameters only and the
    pion-nucleon coupling constants $f^2_0$, $f^2_p$ and $f^2_c$ as
    fitting parameters We take $N=5$ equidistant points with $\Delta r
    = 0.6$fm. $-$ indicates that the corresponding fitting
    $(\lambda_n)^{JS}_{l,l'} =0$. The lowest part of the table shows
    the resulting OPE coupling constants with errors}
  \begin{ruledtabular}  
    \begin{tabular*}{\textwidth}{@{\extracolsep{\fill}}l D{.}{.}{2.6} D{.}{.}{2.6} D{.}{.}{2.6} D{.}{.}{2.6} D{.}{.}{2.6} }
      Wave  & \multicolumn{1}{c}{$\lambda_1$} & 
      \multicolumn{1}{c}{$\lambda_2$} & 
      \multicolumn{1}{c}{$\lambda_3$} & 
      \multicolumn{1}{c}{$\lambda_4$} & 
      \multicolumn{1}{c}{$\lambda_5$}  \\
      \hline\noalign{\smallskip}
      $^1S_{0{\rm np}}$& 1.16(6)               &-0.77(2)               &-0.15(1)               &\multicolumn{1}{c}{$-$}&-0.024(1)              \\
      $^1S_{0{\rm pp}}$& 1.31(2)               &-0.716(5)              &-0.192(2)              &\multicolumn{1}{c}{$-$}&-0.0205(4)             \\
      $^3P_0$          &\multicolumn{1}{c}{$-$}& 0.94(2)               &-0.319(7)              &-0.062(3)              &-0.023(1)              \\
      $^1P_1$          &\multicolumn{1}{c}{$-$}& 1.20(2)               &\multicolumn{1}{c}{$-$}& 0.075(2)              &\multicolumn{1}{c}{$-$}\\
      $^3P_1$          &\multicolumn{1}{c}{$-$}& 1.354(5)              &\multicolumn{1}{c}{$-$}& 0.0570(5)             &\multicolumn{1}{c}{$-$}\\
      $^3S_1$          & 1.79(7)               &-0.47(1)               &\multicolumn{1}{c}{$-$}&-0.072(2)              &\multicolumn{1}{c}{$-$}\\
      $\varepsilon_1$  &\multicolumn{1}{c}{$-$}&-1.65(2)               &-0.33(2)               &-0.233(7)              &-0.018(3)              \\
      $^3D_1$          &\multicolumn{1}{c}{$-$}&\multicolumn{1}{c}{$-$}& 0.40(1)               & 0.070(9)              & 0.021(3)              \\
      $^1D_2$          &\multicolumn{1}{c}{$-$}&-0.20(1)               &-0.206(3)              &\multicolumn{1}{c}{$-$}&-0.0187(3)             \\
      $^3D_2$          &\multicolumn{1}{c}{$-$}&-1.01(3)               &-0.17(2)               &-0.237(6)              &-0.016(2)              \\
      $^3P_2$          &\multicolumn{1}{c}{$-$}&-0.482(1)              &\multicolumn{1}{c}{$-$}&-0.0289(7)             &-0.0037(4)             \\
      $\varepsilon_2$  &\multicolumn{1}{c}{$-$}& 0.32(2)               & 0.190(4)              & 0.050(2)              & 0.0127(6)             \\
      $^3F_2$          &\multicolumn{1}{c}{$-$}& 3.50(6)               &-0.229(5)              &\multicolumn{1}{c}{$-$}&-0.0140(5)              \\
      $^1F_3$          &\multicolumn{1}{c}{$-$}&\multicolumn{1}{c}{$-$}& 0.12(2)               & 0.089(8)              &\multicolumn{1}{c}{$-$}\\
      $^3D_3$          &\multicolumn{1}{c}{$-$}& 0.54(2)               &\multicolumn{1}{c}{$-$}&\multicolumn{1}{c}{$-$}&\multicolumn{1}{c}{$-$}\\
      \hline\noalign{\smallskip}
      \multicolumn{2}{c}{$f^2_{\rm p}$} & \multicolumn{2}{c}{$f^2_0$} & \multicolumn{2}{c}{$f^2_{\rm c}$} \\
      \noalign{\smallskip}
      \hline\noalign{\smallskip}
      \multicolumn{2}{c}{0.0764(4)} & \multicolumn{2}{c}{0.0779(8)} & \multicolumn{2}{c}{0.0758(4)} \\
    \end{tabular*}
  \end{ruledtabular}
\end{table}

\begin{table}[tb]
  \caption{\label{tab:FitParametersPwaves} Same as
    Table~\ref{tab:FitParameters} for a fit with isopsin symmetry
    breaking on the $^1S_0$, $^3P_0$, $^3P_1$ and $^3P_2$ partial
    waves parameters. $^*$ indicates that the np parameter is fixed to
    be the same as the pp parameter}
  \begin{ruledtabular}  
    \begin{tabular*}{\textwidth}{@{\extracolsep{\fill}}l D{.}{.}{2.6} D{.}{.}{2.6} D{.}{.}{2.7} D{.}{.}{2.6} D{.}{.}{2.7} }
      Wave  & \multicolumn{1}{c}{$\lambda_1$} & 
      \multicolumn{1}{c}{$\lambda_2$} & 
      \multicolumn{1}{c}{$\lambda_3$} & 
      \multicolumn{1}{c}{$\lambda_4$} & 
      \multicolumn{1}{c}{$\lambda_5$} \\
      \hline\noalign{\smallskip}
      $^1S_{\rm 0 np}$& 1.07(4)               &-0.708(7)              &-0.192(2)*             &\multicolumn{1}{c}{$-$}&-0.0205(3)*             \\ 
      $^1S_{\rm 0 pp}$& 1.31(2)               &-0.717(5)              &-0.192(2)              &\multicolumn{1}{c}{$-$}&-0.0205(3)              \\ 
      $^3P_{\rm 0 np}$&\multicolumn{1}{c}{$-$}& 0.95(3)               &-0.31(1)               &-0.079(4)              &-0.019(1)               \\ 
      $^3P_{\rm 0 pp}$&\multicolumn{1}{c}{$-$}& 0.94(2)               &-0.319(7)              &-0.063(3)              &-0.022(1)               \\ 
      $^1P_1$         &\multicolumn{1}{c}{$-$}& 1.27(2)               &\multicolumn{1}{c}{$-$}& 0.068(2)              &\multicolumn{1}{c}{$-$} \\ 
      $^3P_{\rm 1 np}$&\multicolumn{1}{c}{$-$}& 1.21(2)               &\multicolumn{1}{c}{$-$}& 0.051(1)              &\multicolumn{1}{c}{$-$} \\ 
      $^3P_{\rm 1 pp}$&\multicolumn{1}{c}{$-$}& 1.364(5)              &\multicolumn{1}{c}{$-$}& 0.0570(6)             &\multicolumn{1}{c}{$-$} \\ 
      $^3S_1$         & 1.54(7)               &-0.39(1)               &\multicolumn{1}{c}{$-$}&-0.071(2)              &\multicolumn{1}{c}{$-$} \\ 
      $\varepsilon_1$ &\multicolumn{1}{c}{$-$}&-1.69(2)               &-0.36(2)               &-0.233(8)              &-0.016(3)               \\ 
      $^3D_1$         &\multicolumn{1}{c}{$-$}&\multicolumn{1}{c}{$-$}& 0.44(2)               & 0.07(1)               & 0.014(3)               \\ 
      $^1D_2$         &\multicolumn{1}{c}{$-$}&-0.19(1)               &-0.207(3)              &\multicolumn{1}{c}{$-$}&-0.0186(3)              \\ 
      $^3D_2$         &\multicolumn{1}{c}{$-$}&-0.97(5)               &-0.21(2)               &-0.234(8)              &-0.016(2)               \\ 
      $^3P_{\rm 2 np}$&\multicolumn{1}{c}{$-$}&-0.445(4)              &\multicolumn{1}{c}{$-$}&-0.043(2)              &-0.0024(7)              \\ 
      $^3P_{\rm 2 pp}$&\multicolumn{1}{c}{$-$}&-0.483(1)              &\multicolumn{1}{c}{$-$}&-0.0282(7)             &-0.0040(4)              \\ 
      $\varepsilon_2$ &\multicolumn{1}{c}{$-$}& 0.30(2)               & 0.191(4)              & 0.051(2)              & 0.0123(6)              \\ 
      $^3F_2$         &\multicolumn{1}{c}{$-$}& 3.41(7)               &-0.222(5)              &\multicolumn{1}{c}{$-$}&-0.0142(6)              \\ 
      $^1F_3$         &\multicolumn{1}{c}{$-$}&\multicolumn{1}{c}{$-$}& 0.23(2)               & 0.061(6)              &\multicolumn{1}{c}{$-$} \\ 
      $^3D_3$         &\multicolumn{1}{c}{$-$}& 0.76(3)               &\multicolumn{1}{c}{$-$}&\multicolumn{1}{c}{$-$}&\multicolumn{1}{c}{$-$} \\ 
      \hline\noalign{\smallskip}
      \multicolumn{2}{c}{$f^2_{\rm p}$} & \multicolumn{2}{c}{$f^2_0$} & \multicolumn{2}{c}{$f^2_{\rm c}$} \\
      \noalign{\smallskip}
      \hline\noalign{\smallskip}
      \multicolumn{2}{c}{0.0761(4)} & \multicolumn{2}{c}{0.0790(9)} & \multicolumn{2}{c}{0.0772(5)} \\
    \end{tabular*}
  \end{ruledtabular}
\end{table}

\subsection{Normality tests}

The standard assumption underlying a conventional $\chi^2$-fit is that
the sum of $\nu$-independent squared gaussian variables belonging to
the normal distribution $N(0,1)$ follows a $\chi^2$ distribution with
$\nu$-degrees of freedom~\cite{evans2004probability}. One can actually
check {\it a posteriori} if the outcoming residuals do indeed fulfill
the initial assumption with a given confidence level. The
self-consistency of the fit is an important test, since it validates
the current statistical analysis, and provides some confidence on the
increase in accuracy that we observed as compared to previous
works. For a number of data much larger than the number of fitting parameters,  
$ N_{\rm Dat} \gg N_{\rm Par}$, the conventional $\chi^2$-test requires 
\begin{eqnarray}
  N_\sigma = \frac{|\chi_{\rm min}^2/\nu-1|}{\sqrt{2/\nu}}    
\end{eqnarray}
with $\nu= N_{\rm Dat}-N_{\rm Par}$ for a $N_\sigma$-standard
deviation confidence level. The tail-sensitive normality test is more
demanding and for the three fits presented on this section are
summarized on Fig.~\ref{fig:qqplots} as rotated quantile-quantile
plots. The tail-sensitive test compares the empirical quantiles of the
residuals with the expected ones from an equally sized sample from the
standard normal distribution. The red bands represent the $95\%$
confidence interval of the normality test. For more details of the
Tail-Sensitive test see~\cite{Perez:2014kpa}.

\subsection{Separate contributions to the fit}

In line with previous studies, it is interesting to decompose the
contributions to the total $\chi^2$ both in terms of the fitted
observables as well as in different energy bins. The separation is
carried out explicitly in Tables~\ref{tab:chi2-obs-pp} and
\ref{tab:chi2-obs-np} for pp and np scattering observables
respectively. As we can see the size of the contributions $\chi^2/N$
are at similar levels for most observables. Note that observables with
a considerable larger or smaller $\chi^2/N$ are also observables with
a small number of data and therefore larger statistical fluctuations
are expected.

Likewise, we can also break up the contributions in order to see the
significance of different energy intervals, see
Table~\ref{tab:bins}. We find that, in agreement with the Nijmegen
analysis (see \cite{Stoks:1993zz,Stoks:1994pi} for comparisons with
previous potentials), there is a relatively large degree of
uniformity in describing data at different energy bins.

\begin{table}[tb]
  \caption{\label{tab:chi2-obs-pp} Contributions to the total $\chi^2$
    for different pp observables. We use the notation of
    \cite{Hoshizaki:1969qt,Bystricky:1976jr}. }
  \begin{tabular*}{\columnwidth}{@{\extracolsep{\fill}}ccccccccccc}
    \toprule
    Observable  & \phantom{11}Code\phantom{11} & $N_{pp}$ & $\chi_{pp}^2$ & $\chi_{pp}^2/N_{pp}$  \\
    \colrule 
    $d\sigma/d\Omega$ & DSG    &   935 &  903.5  &   0.97 \\
    $A_{yy}$          & AYY    &   312 &  339.0  &   1.09 \\
    $D$               & D      &   104 &  135.1  &   1.30 \\
    $P$               & P      &   807 &  832.4  &   1.03 \\
    $A_{zz}$          & AZZ    &    51 &   47.4  &   0.93 \\
    $R$               & R      &   110 &  112.8  &   1.03 \\
    $A$               & A      &    79 &   70.5  &   0.89 \\
    $A_{xx} $         & AXX    &   271 &  250.7  &   0.92 \\
    $C_{kp} $         & CKP    &     2 &    3.1  &   1.57 \\
    $R'$              & RP     &    29 &   11.9  &   0.41 \\
    $M_{s'0sn} $      & MSSN   &    18 &   13.1  &   0.73 \\
    $N_{s'0kn}$       & MSKN   &    18 &    8.5  &   0.47 \\
    $A_{zx}$          & AZX    &   264 &  250.6  &   0.95 \\
    $A'$              & AP     &     6 &    0.8  &   0.14 \\
    \botrule
  \end{tabular*}
\end{table}

\begin{table}
  \caption{\label{tab:chi2-obs-np} Contributions to the total $\chi^2$
    for different np observables. We use the notation of
    \cite{Hoshizaki:1969qt,Bystricky:1976jr} }
  \begin{tabular*}{\columnwidth}{@{\extracolsep{\fill}}ccccccccccc}
    \toprule
    Observable  & \phantom{11}Code\phantom{11} &  $N_{np}$ & $\chi_{np}^2$ & $\chi_{np}^2/N_{np}$  \\
    \colrule 
    $d\sigma/d\Omega$ & DSG    &  1712   & 1803.4 & 1.05 \\
    $D_t$             & DT     &    88   &   83.7 & 0.95 \\
    $A_{yy}$          & AYY    &   119   &   96.0 & 0.81 \\
    $D$               & D      &    29   &   37.1 & 1.28 \\
    $P$               & P      &   977   &  941.7 & 0.96 \\
    $A_{zz}$          & AZZ    &    89   &  108.1 & 1.21 \\
    $R$               & R      &     5   &    4.5 & 0.91 \\
    $R_t$             & RT     &    76   &   72.2 & 0.95 \\
    $R_t'$            & RPT    &     4   &    1.4 & 0.35 \\
    $A_t$             & AT     &    75   &   77.0 & 1.03 \\
    $D_{0s''0k} $     & D0SK   &    29   &   44.0 & 1.52 \\
    $N_{0s''kn}$      & NSKN   &    29   &   25.5 & 0.88 \\
    $N_{0s''sn}$      & NSSN   &    30   &   20.3 & 0.68 \\
    $N_{0nkk}$        & NNKK   &    18   &   13.5 & 0.75 \\
    $A$               & A      &     6   &    2.9 & 0.49 \\
    $\sigma $         & SGT    &   411   &  500.2 & 1.22 \\
    $\Delta \sigma_T $& SGTT   &    20   &   26.3 & 1.31 \\
    $\Delta \sigma_L $& SGTL   &    16   &   18.4 & 1.15 \\
    \botrule
  \end{tabular*}
\end{table}

\begin{table*}
  \caption{\label{tab:bins} The $\chi^{2}$ results of the main
    combined $pp$ and $np$ partial-wave analysis for the 10
    single-energy bins in the range $0 < T_{\rm LAB} < 350 {\rm
      MeV}$.}
  \begin{tabular*}{\textwidth}{@{\extracolsep{\fill}}cccccccccc}
    \toprule
    Bin (MeV)  & $N_{pp}$ & $\chi_{pp}^2$ & $\chi_{pp}^2/N_{pp}$ & $N_{np}$ & $\chi_{np}^2$ & $\chi_{np}^2/N_{np}$ & $N $ & $\chi^2$ & $\chi^2/N$  \\
    \colrule 
    0.0-0.5  & 103 &  107.2  & 1.04 &  46  &  88.2 & 1.92 &  149 &  195.4 & 1.31 \\
    0.5-2    &  82 &   58.8  & 0.72 &  50  &  92.8 & 1.86 &  132 &  151.5 & 1.15 \\
    2-8      &  92 &   80.1  & 0.87 & 122  & 151.0 & 1.24 &  214 &  231.0 & 1.08 \\ 
    8-17     & 124 &  100.3  & 0.81 & 229  & 183.9 & 0.80 &  353 &  284.1 & 0.80 \\
    17-35    & 111 &   85.5  & 0.77 & 346  & 324.2 & 0.94 &  457 &  409.7 & 0.90 \\
    35-75    & 261 &  231.2  & 0.89 & 513  & 559.7 & 1.09 &  774 &  790.9 & 1.02 \\
    75-125   & 152 &  154.8  & 1.02 & 399  & 445.2 & 1.12 &  551 &  600.0 & 1.09 \\
    125-183  & 301 &  300.5  & 1.00 & 372  & 381.7 & 1.03 &  673 &  682.2 & 1.01 \\
    183-290  & 882 &  905.0  & 1.03 & 858  & 841.4 & 0.98 & 1740 & 1746.4 & 1.00 \\
    290-350  & 898 &  956.1  & 1.06 & 798  & 808.1 & 1.01 & 1696 & 1764.1 & 1.04 \\
    \botrule
  \end{tabular*}
\end{table*}

\section{Analysis of Systematic errors}
\label{sec:system}

In this section we seek to identify some sources of systematic
errors. Besides the success of our fits on purely statistical grounds,
it is helpful at this point to analyze why we have chosen our
potential representation and the possible systematic uncertainties
related to it.

\subsection{Anatomy of the potential}
\label{subsec:anatomy}

The present approach uses a coarse grained interaction in the unknown
region, below a cut-off radius $r_c= 3{\rm fm}$. The choice of
$r_c=3$fm, however, is not arbitrary nor blind and in fact it has been
guided by a detailed analysis of existing NN forces. We have checked
that high quality potentials used in the past, are local at large
distances and do implement CD-OPE as the main contribution above 3fm
of strong origin. We remind that a plain extrapolation of the CD-OPE
potential down to the origin presents a short distance $1/r^3$
singularity and a certain regularization is needed which becomes
innocuous at $r > r_c = 3 {\rm fm}$. We have also analyzed quark
models from a cluster viewpoint where there appears a form factor
naturally regulating both electromagnetic Coulomb, OPE and TPE
interactions only below $r_c=1.8-2 {\rm
  fm}$~\cite{Perez:2013cza,Arriola:2016hfi}, so that we can assume
that nucleons interact exchanging one or two pions as point-like
particles for distances larger than $r_c > 1.8 {\rm fm}$. Actually,
this assumption can be validated since lowering down to $r_c=1.2 {\rm
  fm}$ results in large $\chi^2/\nu$ values (see
e.g. \cite{Perez:2014bua,RuizArriola:2016sbf} for a discussion within
chiral perturbation theory).

One good motivation to analyze the NN interaction is the possible
application to nuclear structure calculations. However, the nuclear
many body problem is difficult enough to make specific techniques not
suitable for all representations of the interaction; the form of the
potential matters. Thus, quite often, potentials fitting data are
designed to be suitable for a specific technique. This choice
introduces a bias which acts as a source of systematic errors. In our
previous work~\cite{Perez:2014waa} we have addressed the systematic
uncertainties arising from using several tails and short distance
forms of the potential. The purpose there was to devise a smooth and
non-singular potential in the inner region, friendly for nuclear
structure applications, since it turns out that the delta-shells
produce a long high momentum tail which hinders the nuclear structure
calculations. This includes some bias because, similarly to other
local potentials, smoothness is not a requirement of any physical
significance. Thus, these systematic uncertainties stem from a
prejudice on insisting in a particular form of the potential based on
its possible application in theoretical nuclear physics, and are
relevant within that context.

\subsection{Sampling scale}

The motivation for the coarse grained short distance potential has
been given many times. The sampling scale $\Delta r \sim 1/p_{\rm
  max}$ might be varied from its Nyquist optimal sampling value.  For
a finite range potential that means sampling with more points since
$r_c= n \Delta r$. We generally find that increasing the number of
delta-shells results in over-fitting, i.e., it does not improve the
quality of the fit but it does increase the correlations among the
fitting $\lambda_i$'s parameters, exhibiting a parameter redundancy.
Correlation plots for this optimal sampling situation have been
presented in Ref.~\cite{Perez:2014yla} for the short distance
parameters and in Ref.~\cite{Perez:2014kpa} for the corresponding
counterterms. As it has been discussed in a recent
work~\cite{RuizSimo:2016vsh} the Nyquist sampling works up to LAB
energies as high as 3 GeV.

\subsection{Boundary radius}

In the previous section we have assumed a fixed cut-off radius $r_c=3
{\rm fm}$ above which a CD-OPE potential is assumed. Here we analyze
the robustness of our determination by modifying the cut-off radius,
looking for the cases $r_c=1.8, 2.4, 3.0$ and 3.6 fm. Although the
reasons for choosing $r_c=3 {\rm fm}$ have been explained in
subsection~\ref{subsec:anatomy}, the variation of the cut-off radius
allows to explore the dependence of the statistical analysis on the
particular form of the potential. While this type of cut-off variation
in coordinate space is not entirely equivalent to a cut-off variation
in momentum space, it can provide insight of cut-off dependence in the
latter. Our results are summarized in Table~\ref{tab:fpf0fc-rc}.  For
each value of $r_c$ three PWA are performed. In the first one the
coupling constant $f$ is fixed and not fitted. In the second PWA a
common coupling constant $f$ is fitted as a parameter. In the third
one, the three constants $f_p, f_0$ and $f_c$ are fitted as distinct
parameters.

Several interesting features are worth mentioning. When the short
distance cut-off is shifted towards smaller values the $\chi^2/\nu $
increases several times more than the standard statistical tolerance
$1\pm \sqrt{2/\nu}$.  Larger $\chi^2/\nu$ values generate smaller
uncertainties. This was expected, and it is just a consequence of
the larger penalty to change parameters in a worse fit.

As we see, the best global $\chi^2$ (and nearly equal) values are
obtained for $r_c=3 {\rm fm}$ and $r_c=3.6 {\rm fm}$.  However, we
observe that, in going from $r_c=3.0$ to $r_c=3.6$, the value of
$\chi^2_{pp}$ increases by 40 (with 13 more parameters) whereas the
$\chi^2_{np}$ result decreases by 50.  Increasing the cut-off means
replacing the CD-OPE dependence between 3 and 3.6 by unknown
interactions so that many more partial waves will be charge-dependent,
increasing the number of parameters. At this point the number of CD
parameters becomes rather large. Furthermore, for $r_c=3.6$, the
values obtained for the pion-Nucleon coupling constants are excluded
as unnatural by the Goldberger-Treiman relation shown in
Eq. (\ref{eq:golberger-treiman}).

\subsection{Adding Chiral Potential tails}

The Nijmegen group estimated systematic errors by including different
potential tails, particularly with Heavy Boson Exchange (HBE).  More
recently, the inclusion of charge-independent chiral two-pion exchange
($\chi$TPE) potential~\cite{Kaiser:1997mw}, depending on three
chiral constants, $c_1$, $c_3$, $c_4$, which also appear in $\pi N$
scattering, allowed them to perform a combined fit of $f_p^2$ and
$c_{1,3,4}$ to pp scattering data, obtaining the value
$f_p^2=0.0756(4)$~\cite{Rentmeester:1999vw}, and a simultaneous fit to
pp+np data of a common $f^2$ and $c_{1,3,4}$ \cite{Rentmeester:2003mf}.

In Table~\ref{tab:fpf0fc-TPE} we show several fits of the pion-nucleon
coupling constant $f^2$ after including the $\chi$TPE with different
cut radius $r_c$ on the analysis. In our previous
work~\cite{Perez:2013oba, Perez:2013za, Perez:2014bua} we determined
the value of the chiral constants $c_1$, $c_3$ and $c_4$ from NN data
but maintaining $f$ fixed. The good feature of implementing $\chi$TPE
is that we can generally lower the boundary radius $r_c$ down to the
elementary radius, $r_e=1.8 {\rm fm}$ with a smaller number of
parameters. The outcoming values of the chiral constants should be
compared with the recent re-analysis in $\pi N$ scattering using a
great deal of theoretical constraints~\cite{Siemens:2016jwj}. As with
the case of including only CD-OPE on the potential tail, the
Goldberger-Treiman relation excludes the fits with $r_c = 3.6$ fm. The
unnaturally large values for the chiral constants also calls into
question the analysis with $r_c=3.0$fm and $r_c=2.4$. Finally lowering
the boundary all the way to $r_c=1.2$ fm no longer gives a
satisfactory description of the data, as indicated by the large value
of $\chi^2/\nu$, which is several standard deviations away from the
most likely value.

\begin{table*}[t]
  \caption{\label{tab:fpf0fc-rc} The pion-nucleon coupling constants
    $f_p^ 2$, $f_0^ 2$ and $f_c^ 2$ determined from different fits to
    the Granada-2013 database and their characteristics for the CD-OPE
    potential depending on the cut-off radius $r_c$. Charge dependence
    is only allowed on the $^1S_0$ partial wave.}
  \begin{tabular*}{\textwidth}{@{\extracolsep{\fill}}ccccccccccccc}
    \toprule
    $r_c ({\rm fm})$& $f_p^2$ & $f_0^2$ & $f_c^2$ & $\chi^2_{pp}$ &  $\chi^2_{np}$ & $\chi^2$ &  $N_{\rm Dat}$ & $N_{\rm Par}$   & $\chi^2/\nu $  & $ N_\sigma $ \\ 
    \colrule
    \hline 
    3.6 & 0.075       & idem      & idem      & 3065.13 & 3919.57 & 6984.71 & 6720 & 59 & 1.049 &  2.8 \\ 
    3.6 & 0.0697(3)   & idem      & idem      & 3038.53 & 3913.10 & 6951.63 & 6720 & 60 & 1.044 &  2.5 \\
    3.6 & 0.0689(8)   & 0.085(1)  & 0.0703(8) & 3035.14 & 3897.41 & 6932.55 & 6720 & 62 & 1.041 &  2.4 \\ 
    \hline 
    3.0 & 0.075       & idem      & idem      & 2997.29 & 3957.57 & 6954.86 & 6720 & 46 & 1.042 &  2.4 \\ 
    3.0 & 0.0763(1)   & idem      & idem      & 2995.20 & 3952.85 & 6947.05 & 6720 & 47 & 1.041 &  2.4 \\ 
    3.0 & 0.0764(4)   & 0.0779(8) & 0.0758(4) & 2994.41 & 3950.42 & 6944.83 & 6720 & 49 & 1.041 &  2.4 \\
    \hline 
    2.4 & 0.75        & idem      & idem      & 3120.97 & 4028.61 & 7149.58 & 6718 & 39 & 1.070 &  4.1 \\ 
    2.4 & 0.07568(3)  & idem      & idem      & 3116.56 & 4031.38 & 7147.94 & 6718 & 40 & 1.070 &  4.1 \\ 
    2.4 & 0.0768(3)   & 0.0723(5) & 0.0750(3) & 3115.41 & 4017.76 & 7133.17 & 6718 & 42 & 1.068 &  4.0 \\
    \hline 
    1.8 & 0.75        & idem      & idem      & 4739.51 & 4230.16 & 8969.68 & 6709 & 31 & 1.343 & 19.8 \\ 
    1.8 & 0.076568(5 )& idem      & idem      & 4725.30 & 4212.96 & 8938.26 & 6708 & 32 & 1.339 & 19.6 \\ 
    1.8 & 0.0763(2)   & 0.0786(3) & 0.0765(2) & 4724.73 & 4198.16 & 8922.89 & 6708 & 34 & 1.337 & 19.5 \\
    \botrule
  \end{tabular*}
\end{table*}

\begin{table*}[t]
  \caption{\label{tab:fpf0fc-TPE} The pion-nucleon coupling constant
    $f^2 = f_p^ 2=f_0^ 2=f_c^ 2$ and the chiral constants $c_1$, $c_3$
    and $c_4$ determined from different fits to the Granada-2013
    database and of the CD-OPE plus $\chi TPE$ depending on the
    cut-off radius $r_c$.  Charge dependence is only allowed on the
    $^1S_0$ partial wave.}
  \begin{tabular*}{\textwidth}{@{\extracolsep{\fill}}cccccccccccccc}
    \toprule
    $r_c ({\rm fm})$& $f^2$ & $c_1 ( {\rm GeV}^{-1})$ & $c_3 ( {\rm GeV}^{-1})$ & $c_4 ( {\rm GeV}^{-1}) $ & $\chi^2_{pp}$ &  $\chi^2_{np}$ & $\chi^2$ &  $N_{\rm Dat}$ & $N_{\rm Par}$   & $\chi^2/\nu $  & $ N_\sigma $ \\ 
    \colrule
    \hline 
    3.6 & 0.075      & 1010.0(306) & -990.9(264) &   9.6(140) & 2975.09 & 3879.15 & 6854.24 & 6719 & 63 & 1.030 &  1.7 \\ 
    3.6 & 0.0710(6)  &  978.3(390) & -961.1(353) &  -4.0(148) & 2965.28 & 3869.62 & 6834.90 & 6719 & 64 & 1.027 &  1.6 \\
    \hline  
    3.0 & 0.075      &  -44.4(70)  &   39.5(51)  &  -4.4(26)  & 2979.46 & 3980.27 & 6959.73 & 6721 & 49 & 1.043 &  2.5 \\ 
    3.0 & 0.0763(3)  &  -35.2(79)  &   31.3(60)  &  -6.4(27)  & 2983.95 & 3968.28 & 6952.23 & 6721 & 50 & 1.042 &  2.4 \\ 
    \hline 
    2.4 & 0.075      &  -10.6(18)  &    5.2(10)  &  -2.1(8)   & 3064.38 & 4049.88 & 7114.26 & 6718 & 41 & 1.065 &  3.8 \\ 
    2.4 & 0.0748(2)  &  -11.9(20)  &    6.0(12)  &  -2.3(9)   & 3065.80 & 4048.30 & 7114.11 & 6718 & 42 & 1.066 &  3.8 \\ 
    \hline 
    1.8 & 0.075      &   -1.9(6)   &   -3.7(2)   &   4.4(2)   & 3101.24 & 4059.32 & 7160.56 & 6717 & 33 & 1.071 &  4.1 \\ 
    1.8 & 0.0763(2)  &   -1.6(6)   &   -3.7(3)   &   4.3(2)   & 3077.00 & 4050.22 & 7127.22 & 6717 & 34 & 1.066 &  3.8 \\
    \hline 
    1.2 & 0.075      &  -11.17(9)  &    0.76(2)  &   2.822(2) & 3428.38 & 4659.52 & 8087.90 & 6715 & 25 & 1.209 & 12.1 \\
    1.2 & 0.07500(3) &  -11.17(9)  &    0.76(3)  &   2.821(6) & 3428.28 & 4659.02 & 8087.31 & 6715 & 26 & 1.209 & 12.1 \\
    \botrule
  \end{tabular*}
\end{table*}

\subsection{Sensitivity to particular data}

The selected database provides $3\sigma$ consistent values for the
$\chi^2$ distribution. An important issue concerns the dependence of
our results on the chosen data.  We do not expect all data to
contribute equally to the determination of the coupling constants. In
the past selected data or dedicated experiments have been used to
extract the coupling constant.  Our analysis rests on a global fit,
but it is still interesting to identify the most significant data in
the fit of the coupling constant $f^2$.

From a statistical point of view, this can be done by looking at the
simplest case: the variations $\Delta \chi^2$ due only to variations
on $f$, and by identifying the largest contributions.

\begin{figure}[t]
  \begin{center}
    \epsfig{figure=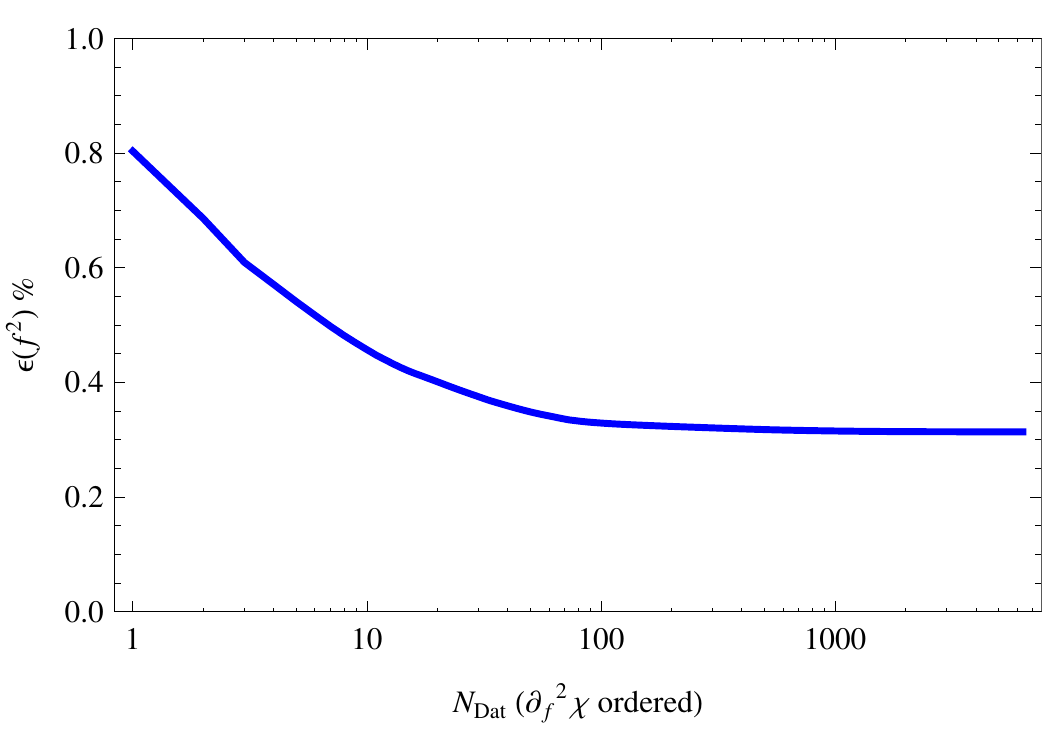,width=8cm,height=6cm}  
  \end{center}
  \caption{Relative error in $f^2$ as a function of the number of data
    {\it ordered} according to a decreasing Hessian value.}
  \label{Fig:hessian}
\end{figure}

The Hessian involving any two fitting parameters $p_i$ and $p_j$
is in general given by 
\begin{eqnarray}
\frac12  \frac{\partial^2
    \chi^2}{\partial p_i \partial p_j} 
\approx \sum_{n=1}^{N_{\rm
      Dat}}\frac{1}{\sigma_n^2} \frac{\partial O_n}{\partial p_i}
  \frac{\partial O_n}{\partial p_j}
\end{eqnarray}
where the standard approximation of neglecting second derivatives has
been made. Here $O_n$ is the $n$th observable in the fit and
$\sigma_n$ is the experimental error. We can look at this sum for one
fitting parameter such as the coupling $f$ {\it after} ordering the
contributions to the Hessian according to their size, i.e. $n \to
\pi(n)$
\begin{eqnarray}
  \frac{1}{\sigma_{\pi(n)}}  \left|\frac{\partial O_{\pi(n)}}{\partial f} \right| >
  \frac{1}{\sigma_{\pi(n-1)}}\left|\frac{\partial O_{\pi(n-1)}}{\partial f} \right|
  %
\end{eqnarray}
and define the error due the first $N$ largest contributions  
\begin{eqnarray}
  \Delta \chi^2_N = \sum_{i=1}^{N} \left[\frac{1}{\sigma_{\pi(n)}} \frac{\partial
      O_{\pi(i)}}{\partial f} \right]^2 (\Delta f)_N^2 \equiv 1
\end{eqnarray}
so that the relative error is $\epsilon_N (f) = \Delta f_N /f $ We
plot in Fig.~\ref{Fig:hessian} the result for $\epsilon_N (f^2 ) = 2
\epsilon_N(f) $ and as we see about 10-20 data build the main
contribution to the precision in $f^2$. These data corresponds to the
deuteron binding energy, the np scattering length, low energy np total
cross sections and low energy pp differential cross sections.

\subsection{Systematics as a function of the number of data}

As already mentioned, the Granada-2013 database is $3\sigma$-self
consistent according to our coarse grained PWA. That implies that we
can treat measurements as independent. On the other hand we expect the
precision will increase with the number of data. Of course, our
selection of data is susceptible to change by gathering more data in
the future. The Cramer-Rao inequality provides a lower bound on the
error on the fitting parameters which can be determined from least
squares
fitting~\cite{evans2004probability}. Thus, errors
will in general be {\it larger} than the $N_{\rm Dat} \to \infty$
case. Given the large amount of data considered in the present
analysis it is of utmost relevance to analyze this point in some more
detail.

\begin{figure}[t]
  \begin{center}
    \epsfig{figure=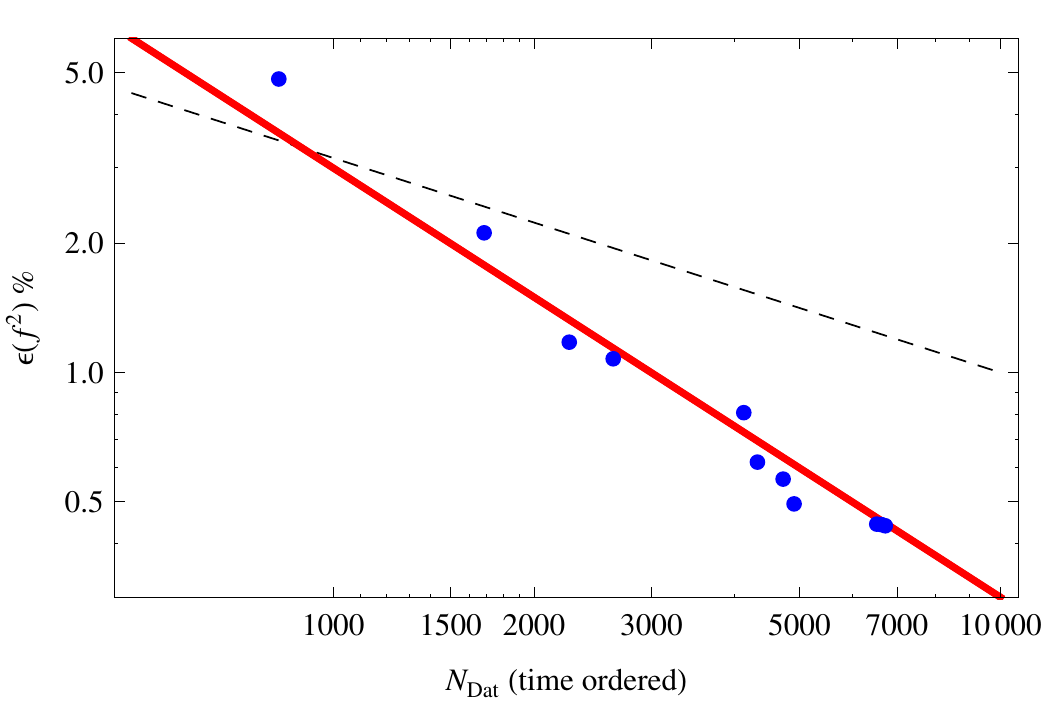,width=8cm,height=6cm}  
  \end{center}
  \caption{Relative error in $f^2$ as a function of the number of data
    {\it ordered} according to their date of publication. Every point
    represents the number of np+pp scattering data extracted from the
    Granada-2013 NN database and starting in 1960 forward in 5 years
    step. We also show the statistical estimate $\epsilon
    (f^2)=1/\sqrt{N_{\rm Par}}$ (black,dashed) and the fitting
    $\epsilon (f^2)=30/N_{\rm Par}$ (red,solid)}
  \label{Fig:relerr}
\end{figure}

Among the many ways of analyzing the systematic uncertainties a
particularly interesting one regards a chronological display of our
self-consistent database as a function of the year where data where
published and hence on the number of scattering data. This can be seen
for pp and pp+np analysis separately in tables \ref{tab:fp-year} and
\ref{tab:f-year} respectively in 5 years intervals. As expected,
accuracy improves when the total number of data $N_{\rm Dat}$ included
in the analysis are increased. Most remarkable is the fact that,
instead of the purely statistical estimate $\Delta f^2/f^2 \sim
1/\sqrt{N_{\rm Dat}}$, a fit to the actual trend reveals more, $\Delta
f^2/f^2 = 29.3/N_{\rm Dat}$, which is in fact better, see
Fig.~\ref{Fig:relerr}. This may be due to the fact that newer data
tends to be more precise than older data. In fact, while the database
contains more np data than pp data, the pp data have smaller
statistical errors and the corresponding fitting parameters tend to be
better determined.

\begin{table}[t]
  \caption{\label{tab:fp-year} The pion-proton-proton coupling
    constant $f_p^ 2$ determined from different fits to the
    Granada-2013 database including only pp data up to a given year.}
  \begin{tabular*}{\columnwidth}{@{\extracolsep{\fill}}c|ccccc}
    \toprule
    \text{Year} & $f_p^2$ & $\Delta f_p^2 $ & $\chi_{pp}^2 $ & $N_{pp}$ &
    $\chi_{pp}^2 / N_{pp}$ \\
    \colrule
    \hline
    1960 & 0.07867 & 0.00421 &  459.50 &  535 & 0.86 \\
    1965 & 0.07568 & 0.00210 &  669.05 &  748 & 0.89 \\
    1970 & 0.07273 & 0.00094 &  978.78 & 1137 & 0.86 \\
    1975 & 0.07317 & 0.00089 & 1149.63 & 1247 & 0.92 \\
    1980 & 0.07339 & 0.00069 & 1486.35 & 1585 & 0.94 \\
    1985 & 0.07443 & 0.00052 & 1559.43 & 1648 & 0.95 \\
    1990 & 0.07528 & 0.00050 & 1774.58 & 1831 & 0.97 \\
    1995 & 0.07542 & 0.00049 & 1809.02 & 1872 & 0.97 \\
    2000 & 0.07596 & 0.00043 & 2985.70 & 3003 & 0.99 \\
    \botrule
  \end{tabular*}
\end{table}

\begin{table}[t]
  \caption{\label{tab:f-year} The pion-nucleon-nucleon coupling
    constant $f^ 2$ determined from different fits to the Granada-2013
    database including only data up to a given year.}
  \begin{tabular*}{\columnwidth}{@{\extracolsep{\fill}}c|ccccccc}
    \toprule
    Year & $f^2$ & $\Delta f^2$ & $\chi_{pp}^2$ & $N_{pp}$ &
    $\chi^2_{np}$ & $N_{np}$ & $\chi^2/N  $ \\
    \colrule
    \hline 
    1960 & 0.07860 & 0.00378 &  460.07 &  535 &  186.92 &  233 & 0.84 \\
    1965 & 0.07740 & 0.00192 &  671.34 &  748 &  791.65 &  836 & 0.92 \\
    1970 & 0.07427 & 0.00088 &  982.23 & 1137 &  922.94 &  981 & 0.90 \\
    1975 & 0.07504 & 0.00082 & 1156.39 & 1247 & 1145.81 & 1221 & 0.93 \\
    1980 & 0.07421 & 0.00061 & 1492.55 & 1585 & 2299.10 & 2311 & 0.97 \\
    1985 & 0.07499 & 0.00046 & 1580.77 & 1648 & 2612.23 & 2584 & 0.99 \\
    1990 & 0.07580 & 0.00043 & 1786.61 & 1831 & 2875.34 & 2806 & 1.01 \\
    1995 & 0.07607 & 0.00039 & 1821.38 & 1872 & 3022.34 & 2950 & 1.00 \\
    2000 & 0.07654 & 0.00034 & 2996.49 & 3003 & 3708.46 & 3528 & 1.03 \\
    2005 & 0.07631 & 0.00034 & 2995.27 & 3003 & 3827.69 & 3634 & 1.03 \\
    2013 & 0.07633 & 0.00014 & 2995.20 & 3003 & 3951.86 & 3717 & 1.03 \\
    \botrule
  \end{tabular*}
\end{table}

\subsection{Summary}

The conclusion of all these investigations is that acceptable and
natural fits produce smaller errorbars than the purely statistical
analysis presented in the previous Section. This is probably due to
the optimal sampling of the interaction complying with Nyquist
theorem.

\section{Conclusions}
\label{sec:conclusions}

\begin{figure}[hpt]
  \begin{center}
    \epsfig{figure=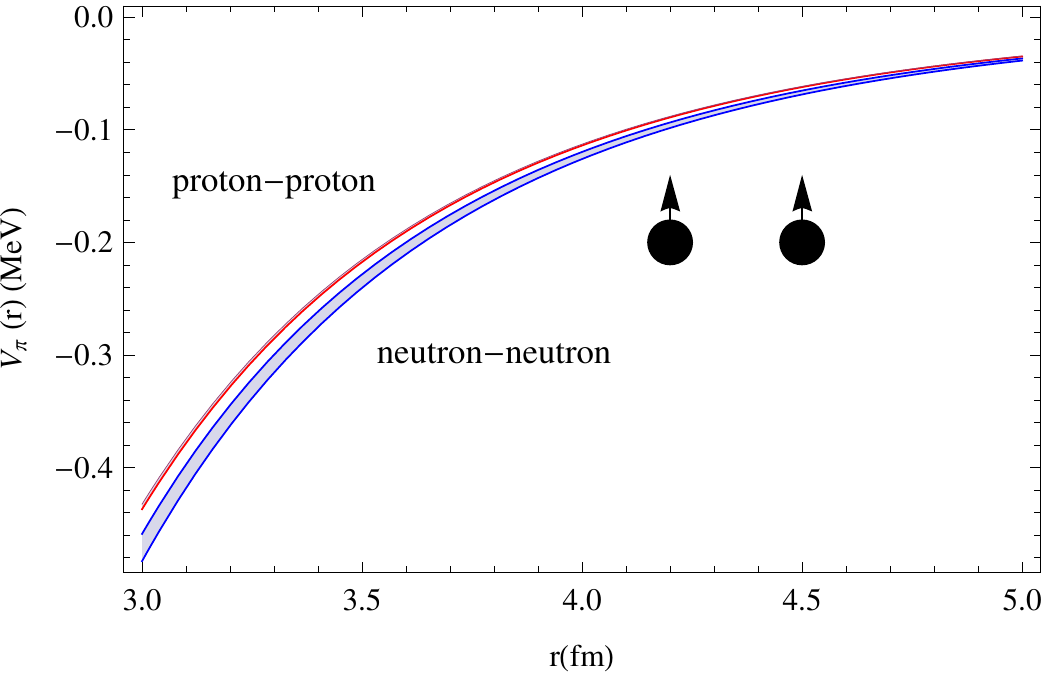,width=4cm,height=4cm}  
    \hskip.3cm
    \epsfig{figure=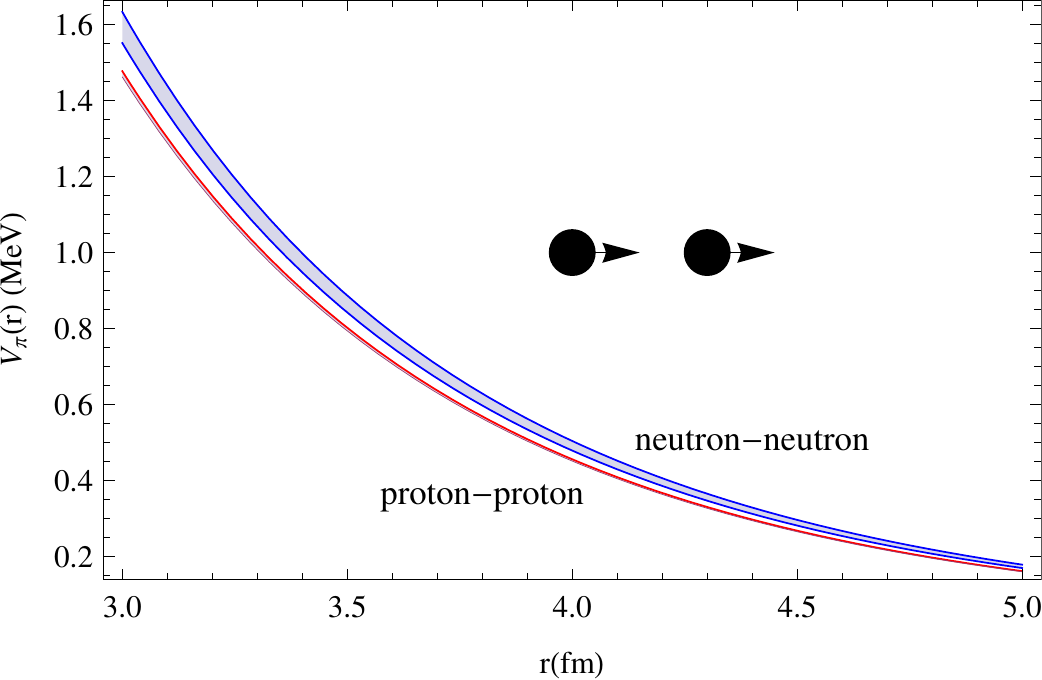,width=4cm,height=4cm} \\
    \vskip.5cm 
    \epsfig{figure=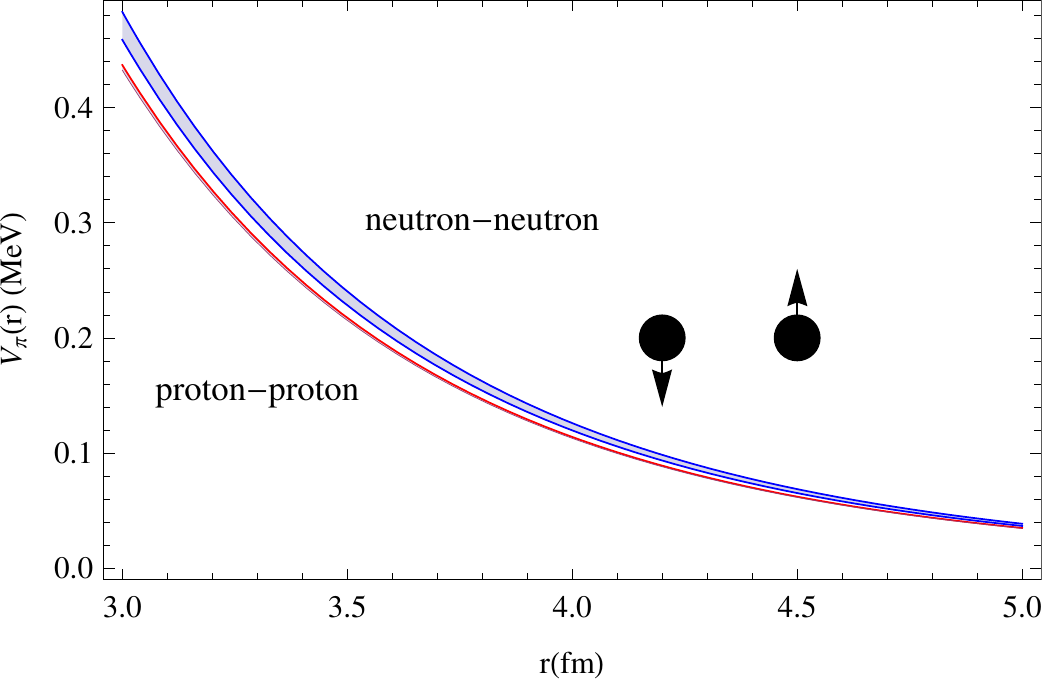,width=4cm,height=4cm}  
    \hskip.3cm
    \epsfig{figure=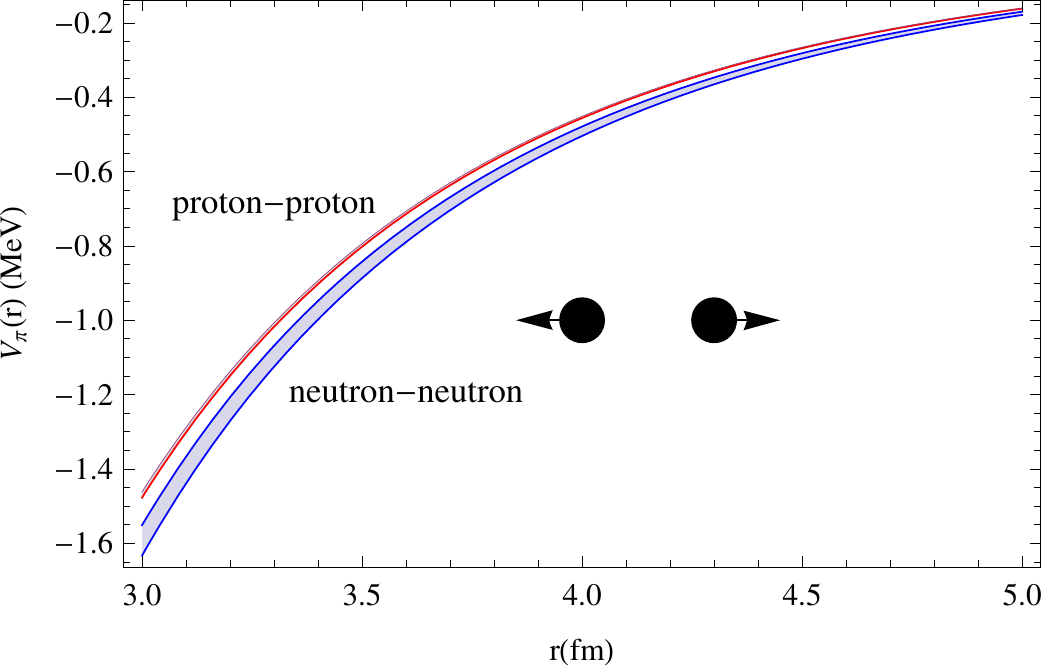,width=4cm,height=4cm}  
  \end{center}
  \caption{Proton-proton and neutron-neutron interaction above 3fm due
    to exchange of a neutral pion for different spin polarization
    states. The bands correspond to the statistical uncertainties from
    a fit to $6713$np+pp scattering data below $T_{\rm LAB}=350 {\rm
      MeV}$ with $\chi^2/\nu=1.039$. }
  \label{Fig:ppnn}
\end{figure}

Since the strong proton-proton and neutron-neutron potentials
correspond to the exchange of a neutral pion, the difference in the
couplings manifests in the difference of the potentials above the
estimated exclusive domain of the CD-OPE interaction. We can
illustrate the main result pictorially in Fig.~\ref{Fig:ppnn} by
choosing the transversely and longitudinally polarized protons and
neutrons.  So we see that in any of the cases considered the strength
of the nn potential is stronger than the pp potential, for instance $
|V_{n \uparrow, n \uparrow}|> |V_{p \uparrow, p \uparrow}|$ for $r >
r_c= 3 {\rm fm}$. Note that we cannot determine the neutron-neutron
interaction below $r_c$, and in particular the corresponding
neutron-neutron scattering length cannot be determined from the
present calculation.

We summarize our points. Using the $3\sigma$ self-consistent
Granada-2013 database for np and pp scattering comprising 
LAB energies below 350 MeV we have investigated
isospin breaking in the pion-nucleon coupling constants by separating
the nuclear potential in two distinct contributions: Above 3 fm we use
charge dependent one pion exchange potential for the strong part along
with electromagnetic and relativistic corrections. Below 3 fm we
regard the interaction as unknown and we coarse grain it down to the
shortest de Broglie wavelength corresponding to pion production
threshold which is about 0.6 fm. With a total number of 55 parameters,
including the three pion-nucleon coupling constants, we describe a
total number of 6741 np and pp data including normalization factors
provided by the experimentalist which a total $\chi^2$ of 6855.5,
which means $\chi^2/\nu=1.025$. We see clear evidence that the
coupling of neutral pions to neutrons is larger than to protons. As a
consequence neutrons interact more strongly than protons.

\begin{acknowledgements}

We thank C. Dominguez, J. Ruiz de Elvira and J.L. Goity for
discussions This work was supported by 6, Spanish Ministerio de
Economia y Competitividad and European FEDER funds (grant
FIS2014-59386-P) and by the Agencia de Innovacion y Desarrollo de
Andalucia (grant No. FQM225).  This work was partly performed under
the auspices of the U.S. Department of Energy by Lawrence Livermore
National Laboratory under Contract No. DE-AC52-07NA27344. Funding was
also provided by the U.S. Department of Energy, Office of Science,
Office of Nuclear Physics under Award No. DE-SC0008511 (NUCLEI SciDAC
Collaboration)

\end{acknowledgements}

\appendix

\section{Operator basis}
\label{sec:operator}
To incorporate charge dependence on $P$ waves two more operators need
to be added to the basis we used previously getting a total of $23$
operators $O^n$. The potential is written as a sum of functions
multiplied by each operator
\begin{equation} 
  V(r) = \sum_{n=1,23} V_n(r) O^n
\end{equation}
The first fourteen operators are charge independent and correspond to
the ones used in the Argonne $v_{14}$ potential
\begin{eqnarray}
  O^{n=1,14} &=& 1, {\tau}_{1}\!\cdot\! {\tau}_{2},\,
  {\sigma}_{1}\!\cdot\! {\sigma}_{2}, ( {\sigma}_{1}\!\cdot\!
  {\sigma}_{2}) ( {\tau}_{1}\!\cdot\! {\tau}_{2}),\, S_{12}, S_{12}(
  {\tau}_{1}\!\cdot\! {\tau}_{2}),\, \nonumber \\ && {\bf
    L}\!\cdot\!{\bf S}, {\bf L}\!\cdot\!{\bf S} ( {\tau}_{1}\!\cdot\!
  {\tau}_{2}), L^{2}, L^{2}( {\tau}_{1}\!\cdot\! {\tau}_{2}),\, L^{2}(
  {\sigma}_{1}\!\cdot\! {\sigma}_{2}), \nonumber\\ & & L^{2}(
  {\sigma}_{1}\!\cdot\! {\sigma}_{2}) ( {\tau}_{1}\!\cdot\!
  {\tau}_{2}),\, ({\bf L}\!\cdot\!{\bf S})^{2}, ({\bf L}\!\cdot\!{\bf
    S})^{2} ( {\tau}_{1}\!\cdot\! {\tau}_{2})\ . \nonumber \\
\end{eqnarray}
These fourteen components are denoted by $c$, $\tau$, $\sigma$,
$\sigma\tau$, $t$, $t\tau$, $ls$, $ls\tau$, $l2$, $l2\tau$,
$l2\sigma$, $l2\sigma\tau$, $ls2$, and $ls2\tau$. The remaining charge
dependent operators are
\begin{eqnarray}
  O^{n=15,21} &=& T_{12}, \, ( {\sigma}_{1}\!\cdot\! {\sigma}_{2})
  T_{12}\, , S_{12}T_{12},\, (\tau_{z1}+\tau_{z2})\ , \nonumber \\ &&(
  {\sigma}_{1}\!\cdot\! {\sigma}_{2}) (\tau_{z1}+\tau_{z2})\ , L^{2}
  T_{12} , L^{2} ( {\sigma}_{1}\!\cdot\! {\sigma}_{2}) T_{12} \, .
  \nonumber \\ && {\bf L}\!\cdot\!{\bf S}T_{12}, ({\bf L}\!\cdot\!{\bf
    S})^{2} T_{12}
\end{eqnarray}
and are labeled as $T$, $\sigma T$,$tT$, $\tau z $,$\sigma \tau z$,
$l2T$, $l2\sigma T$, $ls T$ and $ls2 T$. The first five were
introduced by Wiringa, Stoks and Schiavilla in~\cite{Wiringa:1994wb};
the following two were included in~\cite{Perez:2013jpa} to restrict
the charge dependence to the $^1S_0$ by following certain linear
dependence relations between $V_{T}$, $V_{\sigma T}$, $V_{l2T}$ and
$V_{l2\sigma T}$.  The last two terms are required for the charge
dependence on the $^3P_0$, $^3P_1$ and $^3P_2$ partial waves.

As in our previous analysis we set $V_{tT} = V_{\tau z} = V_{\sigma
  \tau z} = 0$ to exclude charge dependence on the tensor terms and
charge asymmetries.  To restrict the charge dependence to the $S$ and
$P$ waves parameters the remaining potential functions must follow
\begin{eqnarray}
  48 V_{l2 T} &=& -5 V_{T} + 3 V_{\sigma T} + 12 V_{ls T} - 48 V_{ls2 T} \\
  48 V_{\sigma l2 T} &=&  V_{T} - 7 V_{\sigma T} + 4 V_{ls T} - 16 V_{ls2 T} 
\end{eqnarray}

\section{Phase-shifts}
\label{sec:phases} 

Here we provide the pp and np phase-shifts for the lower partial waves
and selected LAB energies with their corresponding errorbars for 
the fit with charge dependence in S and P waves. In the case that
errors are smaller than $10^{-3}$ we just represent it by the symbol
$-$.

\begin{table*}[htb]
 \footnotesize
 \caption{\label{tab:ppIsovectorPS} pp isovector phaseshifts.}
 \begin{ruledtabular}
 \begin{tabular*}{\textwidth}{@{\extracolsep{\fill}} r *{12}{D{.}{.}{3.3}}}
$E_{\rm LAB}$&\multicolumn{1}{c}{$^1S_0$}&\multicolumn{1}{c}{$^1D_2$}&\multicolumn{1}{c}{$^1G_4$}&\multicolumn{1}{c}{$^3P_0$}&\multicolumn{1}{c}{$^3P_1$}&\multicolumn{1}{c}{$^3F_3$}&\multicolumn{1}{c}{$^3P_2$}&\multicolumn{1}{c}{$\epsilon_2$}&\multicolumn{1}{c}{$^3F_2$}&\multicolumn{1}{c}{$^3F_4$}&\multicolumn{1}{c}{$\epsilon_4$}&\multicolumn{1}{c}{$^3H_4$}\\ 
  \hline 
  1 &    32.674 &     0.001 &     0.000 &     0.135 &    -0.081 &    -0.000 &     0.014 &    -0.001 &     0.000 &     0.000 &    -0.000 &     0.000\\
    &\pm 0.003 &    $-$    &    $-$    &    $-$    &    $-$    &    $-$    &    $-$    &    $-$    &    $-$    &    $-$    &    $-$    &    $-$   \\
  5 &    54.836 &     0.043 &     0.000 &     1.599 &    -0.901 &    -0.004 &     0.213 &    -0.052 &     0.002 &     0.000 &    -0.000 &     0.000\\
    &\pm 0.008 &    $-$    &    $-$    &\pm 0.005 &\pm 0.003 &    $-$    &\pm 0.001 &    $-$    &    $-$    &    $-$    &    $-$    &    $-$   \\
 10 &    55.238 &     0.166 &     0.003 &     3.772 &    -2.054 &    -0.032 &     0.648 &    -0.204 &     0.013 &     0.001 &    -0.004 &     0.000\\
    &\pm 0.011 &\pm 0.001 &    $-$    &\pm 0.011 &\pm 0.006 &    $-$    &\pm 0.002 &\pm 0.001 &    $-$    &    $-$    &    $-$    &    $-$   \\
 25 &    48.759 &     0.699 &     0.040 &     8.666 &    -4.888 &    -0.234 &     2.490 &    -0.820 &     0.107 &     0.019 &    -0.050 &     0.004\\
    &\pm 0.014 &\pm 0.002 &    $-$    &\pm 0.027 &\pm 0.010 &\pm 0.001 &\pm 0.006 &\pm 0.003 &\pm 0.001 &    $-$    &    $-$    &    $-$   \\
 50 &    39.133 &     1.711 &     0.154 &    11.577 &    -8.224 &    -0.696 &     5.856 &    -1.719 &     0.346 &     0.104 &    -0.200 &     0.027\\
    &\pm 0.018 &\pm 0.004 &\pm 0.001 &\pm 0.046 &\pm 0.013 &\pm 0.004 &\pm 0.011 &\pm 0.005 &\pm 0.002 &\pm 0.001 &\pm 0.001 &    $-$   \\
100 &    25.340 &     3.774 &     0.422 &     9.535 &   -13.260 &    -1.502 &    10.986 &    -2.650 &     0.842 &     0.466 &    -0.561 &     0.112\\
    &\pm 0.036 &\pm 0.009 &\pm 0.002 &\pm 0.072 &\pm 0.021 &\pm 0.010 &\pm 0.024 &\pm 0.009 &\pm 0.009 &\pm 0.005 &\pm 0.002 &\pm 0.001\\
150 &    15.116 &     5.618 &     0.703 &     4.839 &   -17.637 &    -2.077 &    14.021 &    -2.921 &     1.203 &     1.019 &    -0.883 &     0.225\\
    &\pm 0.050 &\pm 0.015 &\pm 0.006 &\pm 0.079 &\pm 0.027 &\pm 0.020 &\pm 0.023 &\pm 0.013 &\pm 0.017 &\pm 0.010 &\pm 0.003 &\pm 0.002\\
200 &     6.892 &     7.206 &     1.000 &    -0.214 &   -21.554 &    -2.471 &    15.803 &    -2.907 &     1.334 &     1.638 &    -1.133 &     0.351\\
    &\pm 0.060 &\pm 0.022 &\pm 0.012 &\pm 0.071 &\pm 0.039 &\pm 0.037 &\pm 0.027 &\pm 0.017 &\pm 0.023 &\pm 0.017 &\pm 0.005 &\pm 0.006\\
250 &     0.178 &     8.552 &     1.295 &    -5.098 &   -24.984 &    -2.623 &    16.739 &    -2.698 &     1.224 &     2.205 &    -1.311 &     0.479\\
    &\pm 0.075 &\pm 0.025 &\pm 0.017 &\pm 0.068 &\pm 0.055 &\pm 0.054 &\pm 0.034 &\pm 0.023 &\pm 0.029 &\pm 0.024 &\pm 0.006 &\pm 0.013\\
300 &    -5.222 &     9.571 &     1.571 &    -9.601 &   -27.919 &    -2.418 &    16.981 &    -2.293 &     0.918 &     2.659 &    -1.439 &     0.591\\
    &\pm 0.102 &\pm 0.032 &\pm 0.019 &\pm 0.095 &\pm 0.070 &\pm 0.066 &\pm 0.034 &\pm 0.032 &\pm 0.041 &\pm 0.029 &\pm 0.008 &\pm 0.020\\
350 &    -9.447 &    10.140 &     1.832 &   -13.545 &   -30.348 &    -1.820 &    16.635 &    -1.707 &     0.454 &     3.012 &    -1.552 &     0.670\\
    &\pm 0.138 &\pm 0.055 &\pm 0.027 &\pm 0.152 &\pm 0.082 &\pm 0.080 &\pm 0.031 &\pm 0.042 &\pm 0.058 &\pm 0.047 &\pm 0.010 &\pm 0.027
 \end{tabular*}
 \end{ruledtabular}
\end{table*}

\begin{table*}[htb]
 \footnotesize
 \caption{\label{tab:npIsovectorPS}np isovector phaseshifts.}
 \begin{ruledtabular}
  \begin{tabular*}{\textwidth}{@{\extracolsep{\fill}} r *{12}{D{.}{.}{3.3}}}
$E_{\rm LAB}$&\multicolumn{1}{c}{$^1S_0$}&\multicolumn{1}{c}{$^1D_2$}&\multicolumn{1}{c}{$^1G_4$}&\multicolumn{1}{c}{$^3P_0$}&\multicolumn{1}{c}{$^3P_1$}&\multicolumn{1}{c}{$^3F_3$}&\multicolumn{1}{c}{$^3P_2$}&\multicolumn{1}{c}{$\epsilon_2$}&\multicolumn{1}{c}{$^3F_2$}&\multicolumn{1}{c}{$^3F_4$}&\multicolumn{1}{c}{$\epsilon_4$}&\multicolumn{1}{c}{$^3H_4$}\\ 
  1 &    62.047 &     0.001 &     0.000 &     0.183 &    -0.105 &    -0.000 &     0.025 &    -0.001 &     0.000 &     0.000 &    -0.000 &     0.000\\
    &\pm 0.024 &    $-$    &    $-$    &\pm 0.003 &\pm 0.002 &    $-$    &    $-$    &    $-$    &    $-$    &    $-$    &    $-$    &    $-$   \\
  5 &    63.559 &     0.041 &     0.000 &     1.683 &    -0.911 &    -0.004 &     0.284 &    -0.049 &     0.002 &     0.000 &    -0.000 &     0.000\\
    &\pm 0.046 &\pm 0.001 &    $-$    &\pm 0.029 &\pm 0.016 &    $-$    &\pm 0.004 &\pm 0.001 &    $-$    &    $-$    &    $-$    &    $-$   \\
 10 &    59.851 &     0.155 &     0.002 &     3.823 &    -1.996 &    -0.026 &     0.796 &    -0.185 &     0.011 &     0.001 &    -0.003 &     0.000\\
    &\pm 0.053 &\pm 0.003 &    $-$    &\pm 0.058 &\pm 0.031 &\pm 0.001 &\pm 0.009 &\pm 0.004 &    $-$    &    $-$    &    $-$    &    $-$   \\
 25 &    50.712 &     0.673 &     0.032 &     8.699 &    -4.666 &    -0.195 &     2.846 &    -0.765 &     0.091 &     0.017 &    -0.040 &     0.003\\
    &\pm 0.062 &\pm 0.012 &\pm 0.001 &\pm 0.096 &\pm 0.053 &\pm 0.005 &\pm 0.027 &\pm 0.014 &\pm 0.002 &\pm 0.001 &\pm 0.001 &    $-$   \\
 50 &    40.225 &     1.695 &     0.134 &    11.682 &    -7.822 &    -0.594 &     6.364 &    -1.660 &     0.309 &     0.105 &    -0.170 &     0.021\\
    &\pm 0.083 &\pm 0.019 &\pm 0.003 &\pm 0.109 &\pm 0.059 &\pm 0.015 &\pm 0.045 &\pm 0.020 &\pm 0.006 &\pm 0.004 &\pm 0.004 &\pm 0.001\\
100 &    26.129 &     3.760 &     0.391 &     9.484 &   -12.592 &    -1.299 &    11.070 &    -2.658 &     0.780 &     0.514 &    -0.507 &     0.095\\
    &\pm 0.138 &\pm 0.018 &\pm 0.009 &\pm 0.163 &\pm 0.080 &\pm 0.036 &\pm 0.061 &\pm 0.009 &\pm 0.013 &\pm 0.021 &\pm 0.011 &\pm 0.002\\
150 &    15.917 &     5.568 &     0.671 &     4.399 &   -16.716 &    -1.915 &    13.334 &    -3.000 &     1.117 &     1.130 &    -0.828 &     0.204\\
    &\pm 0.191 &\pm 0.017 &\pm 0.013 &\pm 0.216 &\pm 0.120 &\pm 0.062 &\pm 0.082 &\pm 0.020 &\pm 0.020 &\pm 0.039 &\pm 0.015 &\pm 0.005\\
200 &     7.810 &     7.118 &     0.961 &    -1.038 &   -20.374 &    -2.536 &    14.510 &    -3.028 &     1.217 &     1.773 &    -1.091 &     0.334\\
    &\pm 0.243 &\pm 0.023 &\pm 0.017 &\pm 0.261 &\pm 0.170 &\pm 0.085 &\pm 0.109 &\pm 0.027 &\pm 0.024 &\pm 0.054 &\pm 0.014 &\pm 0.011\\
250 &     1.289 &     8.435 &     1.238 &    -6.196 &   -23.543 &    -2.989 &    15.158 &    -2.834 &     1.072 &     2.318 &    -1.290 &     0.464\\
    &\pm 0.302 &\pm 0.025 &\pm 0.020 &\pm 0.304 &\pm 0.224 &\pm 0.101 &\pm 0.125 &\pm 0.030 &\pm 0.030 &\pm 0.070 &\pm 0.010 &\pm 0.018\\
300 &    -3.856 &     9.421 &     1.488 &   -10.830 &   -26.215 &    -2.958 &    15.386 &    -2.426 &     0.733 &     2.732 &    -1.441 &     0.561\\
    &\pm 0.367 &\pm 0.033 &\pm 0.021 &\pm 0.356 &\pm 0.274 &\pm 0.135 &\pm 0.122 &\pm 0.036 &\pm 0.042 &\pm 0.090 &\pm 0.008 &\pm 0.027\\
350 &    -7.791 &     9.947 &     1.720 &   -14.770 &   -28.375 &    -2.224 &    15.185 &    -1.828 &     0.244 &     3.053 &    -1.574 &     0.593\\
    &\pm 0.439 &\pm 0.057 &\pm 0.028 &\pm 0.427 &\pm 0.318 &\pm 0.250 &\pm 0.107 &\pm 0.045 &\pm 0.059 &\pm 0.126 &\pm 0.011 &\pm 0.036
 \end{tabular*}
 \end{ruledtabular}
\end{table*}

\begin{table*}
 \footnotesize
 \caption{\label{tab:npIsoscalarPS}np isoscalar phaseshifts.}
 \begin{ruledtabular}
 \begin{tabular*}{\textwidth}{@{\extracolsep{\fill}} r *{12}{D{.}{.}{3.3}}}
 $E_{\rm LAB}$&\multicolumn{1}{c}{$^1P_1$}&\multicolumn{1}{c}{$^1F_3$}&\multicolumn{1}{c}{$^3D_2$}&\multicolumn{1}{c}{$^3G_4$}&\multicolumn{1}{c}{$^3S_1$}&\multicolumn{1}{c}{$\epsilon_1$}&\multicolumn{1}{c}{$^3D_1$}&\multicolumn{1}{c}{$^3D_3$}&\multicolumn{1}{c}{$\epsilon_3$}&\multicolumn{1}{c}{$^3G_3$}\\ 
  \hline 
  1 &    -0.191 &    -0.000 &     0.006 &     0.000 &   147.685 &     0.105 &    -0.005 &     0.000 &     0.000 &    -0.000\\
    &    $-$    &    $-$    &    $-$    &    $-$    &\pm 0.017 &\pm 0.001 &    $-$    &    $-$    &    $-$    &    $-$   \\
  5 &    -1.528 &    -0.010 &     0.226 &     0.001 &   118.043 &     0.654 &    -0.186 &     0.002 &     0.013 &    -0.000\\
    &\pm 0.003 &    $-$    &    $-$    &    $-$    &\pm 0.024 &\pm 0.005 &    $-$    &    $-$    &    $-$    &    $-$   \\
 10 &    -3.119 &    -0.066 &     0.876 &     0.012 &   102.425 &     1.112 &    -0.690 &     0.005 &     0.083 &    -0.003\\
    &\pm 0.009 &    $-$    &\pm 0.001 &    $-$    &\pm 0.034 &\pm 0.011 &\pm 0.002 &    $-$    &    $-$    &    $-$   \\
 25 &    -6.413 &    -0.435 &     3.839 &     0.177 &    80.364 &     1.696 &    -2.843 &     0.039 &     0.572 &    -0.055\\
    &\pm 0.029 &    $-$    &\pm 0.010 &    $-$    &\pm 0.059 &\pm 0.026 &\pm 0.009 &\pm 0.001 &    $-$    &    $-$   \\
 50 &    -9.656 &    -1.173 &     9.265 &     0.755 &    62.489 &     2.032 &    -6.496 &     0.292 &     1.658 &    -0.274\\
    &\pm 0.062 &\pm 0.002 &\pm 0.037 &\pm 0.001 &\pm 0.078 &\pm 0.047 &\pm 0.026 &\pm 0.005 &\pm 0.004 &\pm 0.001\\
100 &   -14.214 &    -2.304 &    17.776 &     2.321 &    43.135 &     2.515 &   -12.295 &     1.321 &     3.532 &    -1.007\\
    &\pm 0.097 &\pm 0.015 &\pm 0.076 &\pm 0.014 &\pm 0.086 &\pm 0.082 &\pm 0.054 &\pm 0.020 &\pm 0.018 &\pm 0.011\\
150 &   -18.203 &    -3.097 &    22.620 &     3.986 &    30.862 &     2.927 &   -16.683 &     2.338 &     4.833 &    -1.873\\
    &\pm 0.114 &\pm 0.036 &\pm 0.096 &\pm 0.045 &\pm 0.098 &\pm 0.110 &\pm 0.081 &\pm 0.042 &\pm 0.031 &\pm 0.036\\
200 &   -21.765 &    -3.831 &    24.866 &     5.583 &    21.462 &     3.272 &   -20.284 &     2.931 &     5.790 &    -2.718\\
    &\pm 0.142 &\pm 0.053 &\pm 0.123 &\pm 0.080 &\pm 0.123 &\pm 0.131 &\pm 0.114 &\pm 0.067 &\pm 0.040 &\pm 0.064\\
250 &   -24.856 &    -4.634 &    25.699 &     7.011 &    13.557 &     3.677 &   -23.290 &     3.067 &     6.565 &    -3.516\\
    &\pm 0.181 &\pm 0.061 &\pm 0.149 &\pm 0.102 &\pm 0.148 &\pm 0.142 &\pm 0.130 &\pm 0.095 &\pm 0.055 &\pm 0.083\\
300 &   -27.487 &    -5.498 &    25.991 &     8.197 &     6.566 &     4.262 &   -25.639 &     2.870 &     7.195 &    -4.314\\
    &\pm 0.221 &\pm 0.077 &\pm 0.194 &\pm 0.117 &\pm 0.171 &\pm 0.175 &\pm 0.143 &\pm 0.129 &\pm 0.069 &\pm 0.101\\
350 &   -29.666 &    -6.335 &    26.295 &     9.101 &     0.223 &     5.068 &   -27.156 &     2.492 &     7.647 &    -5.173\\
    &\pm 0.257 &\pm 0.129 &\pm 0.312 &\pm 0.142 &\pm 0.193 &\pm 0.265 &\pm 0.210 &\pm 0.166 &\pm 0.082 &\pm 0.146
 \end{tabular*}
 \end{ruledtabular}
\end{table*}


\begin{thebibliography}{10}

\bibitem{Yukawa:1935xg}
H. Yukawa,
\newblock Proc. Phys. Math. Soc. Jap. 17 (1935) 48,
\newblock [Prog. Theor. Phys. Suppl.1,1(1935)].

\bibitem{pauli1948meson}
W. Pauli,
\newblock Meson theory of nuclear forces (Interscience Publishers, 1948).

\bibitem{Dumbrajs:1983jd}
O. Dumbrajs et~al.,
\newblock Nucl.Phys. B216 (1983) 277.

\bibitem{deSwart:1997ep}
J. de~Swart, M. Rentmeester and R. Timmermans,
\newblock PiN Newslett. 13 (1997) 96, nucl-th/9802084.

\bibitem{Sainio:1999ba}
M. Sainio,
\newblock PiN Newslett. 15 (1999) 156, hep-ph/9912337.

\bibitem{Bugg:2004cm}
D. Bugg,
\newblock Eur.Phys.J. C33 (2004) 505.

\bibitem{PhysRev.57.260}
H.A. Bethe,
\newblock Phys. Rev. 57 (1940) 260.

\bibitem{Bethe:1940zz}
H. Bethe,
\newblock Phys.Rev. 57 (1940) 390.

\bibitem{Goldberger:1958tr}
M.L. Goldberger and S.B. Treiman,
\newblock Phys. Rev. 110 (1958) 1178.

\bibitem{Nambu:1960xd}
Y. Nambu,
\newblock Phys. Rev. Lett. 4 (1960) 380.

\bibitem{Chew:1958zz}
G.F. Chew,
\newblock Phys. Rev. 112 (1958) 1380.

\bibitem{Cziffra:1959zza}
P. Cziffra et~al.,
\newblock Phys. Rev. 114 (1959) 880.

\bibitem{MacGregor:1959zz}
M.H. MacGregor, M.J. Moravcsik and H.P. Stapp,
\newblock Phys. Rev. 116 (1959) 1248.

\bibitem{Signell:1960zz}
P.S. Signell,
\newblock Phys. Rev. Lett. 5 (1960) 474.

\bibitem{macgregor1964determination}
M. MacGregor, R. Arndt and A. Dubow,
\newblock Physical Review 135 (1964) B628.

\bibitem{Bugg:1973rv}
D.V. Bugg, A.A. Carter and J.R. Carter,
\newblock Phys. Lett. B44 (1973) 278.

\bibitem{Arndt:1990cn}
R.A. Arndt et~al.,
\newblock Phys. Rev. Lett. 65 (1990) 157.

\bibitem{Arndt:1994bu}
R. Arndt, R. Workman and M. Pavan,
\newblock Phys.Rev. C49 (1994) 2729.

\bibitem{Ericson:2000md}
T.E.O. Ericson, B. Loiseau and A.W. Thomas,
\newblock Phys.Rev. C66 (2002) 014005, hep-ph/0009312.

\bibitem{Arndt:2006bf}
R.A. Arndt et~al.,
\newblock Phys. Rev. C74 (2006) 045205, nucl-th/0605082.

\bibitem{Baru:2010xn}
V. Baru et~al.,
\newblock Phys.Lett. B694 (2011) 473, 1003.4444.

\bibitem{Baru:2011bw}
V. Baru et~al.,
\newblock Nucl.Phys. A872 (2011) 69, 1107.5509.

\bibitem{Timmermans:1990tz}
R.G.E. Timmermans, T.A. Rijken and J.J. de~Swart,
\newblock Phys. Rev. Lett. 67 (1991) 1074.

\bibitem{Stoks:1993tb}
V. Stoks et~al.,
\newblock Phys.Rev. C48 (1993) 792.

\bibitem{Klomp:1991vz}
R. Klomp, V. Stoks and J. de~Swart,
\newblock Phys.Rev. C44 (1991) 1258.

\bibitem{Stoks:1992ja}
V.G. Stoks, R. Timmermans and J. de~Swart,
\newblock Phys.Rev. C47 (1993) 512, nucl-th/9211007.

\bibitem{Kaiser:1997mw}
N. Kaiser, R. Brockmann and W. Weise,
\newblock Nucl. Phys. A625 (1997) 758, nucl-th/9706045.

\bibitem{Rentmeester:1999vw}
M.C.M. Rentmeester et~al.,
\newblock Phys. Rev. Lett. 82 (1999) 4992, nucl-th/9901054.

\bibitem{Rentmeester:2003mf}
M.C.M. Rentmeester, R.G.E. Timmermans and J.J. de~Swart,
\newblock Phys. Rev. C67 (2003) 044001, nucl-th/0302080.

\bibitem{evans2004probability}
M.J. Evans and J.S. Rosenthal,
\newblock Probability and statistics: The science of uncertainty (Macmillan,
  2004).

\bibitem{Dobaczewski:2014jga}
J. Dobaczewski, W. Nazarewicz and P.G. Reinhard,
\newblock J. Phys. G41 (2014) 074001, 1402.4657.

\bibitem{Bergervoet:1988zz}
J. Bergervoet et~al.,
\newblock Phys.Rev. C38 (1988) 15.

\bibitem{Perez:2013cza}
R. Navarro~Pérez, J.E. Amaro and E. Ruiz~Arriola,
\newblock Few Body Syst. 55 (2014) 983, 1310.8167.

\bibitem{Perez:2013jpa}
R. Navarro~Pérez, J.E. Amaro and E. Ruiz~Arriola,
\newblock Phys. Rev. C88 (2013) 064002, 1310.2536,
\newblock [Erratum: Phys. Rev.C91,no.2,029901(2015)].

\bibitem{Perez:2014yla}
R. Navarro~Pérez, J.E. Amaro and E. Ruiz~Arriola,
\newblock Phys. Rev. C89 (2014) 064006, 1404.0314.

\bibitem{Perez:2014kpa}
R. Navarro~Pérez, J.E. Amaro and E. Ruiz~Arriola,
\newblock J. Phys. G42 (2015) 034013, 1406.0625.

\bibitem{Aldor2013}
S. Aldor-Noiman et~al.,
\newblock Amer. Statist. 67 (2013) 249.

\bibitem{Perez:2015pea}
R. Navarro~Pérez, E. Ruiz~Arriola and J. Ruiz~de Elvira,
\newblock Phys. Rev. D91 (2015) 074014, 1502.03361.

\bibitem{Ericson:1995gr}
T.E.O. Ericson et~al.,
\newblock Phys. Rev. Lett. 75 (1995) 1046.

\bibitem{Rahm:1998jt}
J. Rahm et~al.,
\newblock Phys. Rev. C57 (1998) 1077.

\bibitem{Weinberg:1996kr}
S. Weinberg,
\newblock {The quantum theory of fields. Vol. 2: Modern applications}
  (Cambridge University Press, 2013).

\bibitem{Rentmeester:1998vf}
M.C.M. Rentmeester, R.A.M. Klomp and J.J. de~Swart,
\newblock Phys. Rev. Lett. 81 (1998) 5253, nucl-th/9812020.

\bibitem{Ericson:1998fq}
T.E.O. Ericson et~al.,
\newblock Phys. Rev. Lett. 81 (1998) 5254.

\bibitem{Blomgren:2000wq}
J. Blomgren, editor,
\newblock {Critical issues in the determination of the pion nucleon coupling
  constant. Proceedings, Workshop, Uppsala, Sweden, June 7-8, 1999} Vol. T87,
  2000.

\bibitem{Sarsour:2004xx}
M. Sarsour et~al.,
\newblock Phys. Rev. Lett. 94 (2005) 082303, nucl-ex/0412026.

\bibitem{Sarsour:2006fd}
M. Sarsour et~al.,
\newblock Phys. Rev. C74 (2006) 044003, nucl-ex/0602017.

\bibitem{Aoki:2011ep}
Sinya AOKI for HAL QCD Collaboration, S. Aoki,
\newblock Prog.Part.Nucl.Phys. 66 (2011) 687, 1107.1284.

\bibitem{Aoki:2013tba}
S. Aoki,
\newblock Eur.Phys.J. A49 (2013) 81, 1309.4150.

\bibitem{Aoki:2009ji}
S. Aoki, T. Hatsuda and N. Ishii,
\newblock Prog.Theor.Phys. 123 (2010) 89, 0909.5585.

\bibitem{Aviles:1973ee}
J.B. Aviles,
\newblock Phys. Rev. C6 (1972) 1467.

\bibitem{Perez:2016vzj}
R.N. Perez, J.E. Amaro and E. Ruiz~Arriola,
\newblock Int. J. Mod. Phys. E25 (2016) 1641009, 1601.08220.

\bibitem{Dominguez:1984ka}
C.A. Dominguez,
\newblock Riv. Nuovo Cim. 8N6 (1985) 1.

\bibitem{Goity:1999by}
J.L. Goity et~al.,
\newblock Phys.Lett. B454 (1999) 115, hep-ph/9901374.

\bibitem{Goity:2002uh}
J. Goity and J. Saez,
\newblock report JLAB-THY-03-26 (2002).


\bibitem{Bernard:1996gq}
V. Bernard, N. Kaiser and U.G. Meissner,
\newblock Nucl. Phys. A615 (1997) 483, hep-ph/9611253.

\bibitem{Ledwig:2014cla}
T. Ledwig et~al.,
\newblock Phys. Rev. D90 (2014) 114020, 1407.3750.

\bibitem{Thomas:1989tv}
A.W. Thomas and K. Holinde,
\newblock Phys.Rev.Lett. 63 (1989) 2025.

\bibitem{Braun:2008eh}
R. Braun et~al.,
\newblock Phys.Lett. B660 (2008) 161, 0801.4600.

\bibitem{Gross:2008pd}
F. Gross and A. Stadler,
\newblock Phys.Lett. B668 (2008) 163, 0808.2962.

\bibitem{Weisel:2010zz}
G. Weisel, R. Braun and W. Tornow,
\newblock Phys.Rev. C82 (2010) 027001.

\bibitem{Wiringa:1994wb}
R.B. Wiringa, V. Stoks and R. Schiavilla,
\newblock Phys.Rev. C51 (1995) 38, nucl-th/9408016.

\bibitem{Machleidt:2000ge}
R. Machleidt,
\newblock Phys.Rev. C63 (2001) 024001, nucl-th/0006014.

\bibitem{Perez:2013oba}
R. Navarro~Pérez, J.E. Amaro and E. Ruiz~Arriola,
\newblock Phys. Rev. C89 (2014) 024004, 1310.6972.

\bibitem{Perez:2013mwa}
R. Navarro~Pérez, J.E. Amaro and E. Ruiz~Arriola,
\newblock Phys. Rev. C88 (2013) 024002, 1304.0895,
\newblock [Erratum: Phys. Rev.C88,no.6,069902(2013)].

\bibitem{Stoks:1993zz}
V. Stoks and J.J. de~Swart,
\newblock Phys. Rev. C47 (1993) 761.

\bibitem{Stoks:1994pi}
V.G.J. Stoks and J.J. de~Swart,
\newblock Phys. Rev. C52 (1995) 1698, nucl-th/9411002.

\bibitem{Hoshizaki:1969qt}
N. Hoshizaki,
\newblock Prog. Theor. Phys. Suppl. 42 (1969) 107.

\bibitem{Bystricky:1976jr}
J. Bystricky, F. Lehar and P. Winternitz,
\newblock J. Phys.(France) 39 (1978) 1.

\bibitem{Arriola:2016hfi}
E. Ruiz~Arriola, J.E. Amaro and R. Navarro~Pérez,
\newblock Mod. Phys. Lett. A31 (2016) 1630027, 1606.02171.

\bibitem{Perez:2014bua}
R. Navarro~Pérez, J.E. Amaro and E. Ruiz~Arriola,
\newblock Phys. Rev. C91 (2015) 054002, 1411.1212.

\bibitem{RuizArriola:2016sbf}
E. Ruiz~Arriola, J.E. Amaro and R. Navarro~Perez,
\newblock EPJ Web Conf. 137 (2017) 09006, 1611.02607.

\bibitem{Perez:2014waa}
R. Navarro~Pérez, J.E. Amaro and E. Ruiz~Arriola,
\newblock J. Phys. G43 (2016) 114001, 1410.8097.

\bibitem{RuizSimo:2016vsh}
I. Ruiz~Simo et~al.,
\newblock (2016), 1612.06228.

\bibitem{Perez:2013za}
R. Navarro~Pérez, J.E. Amaro and E. Ruiz~Arriola,
\newblock PoS CD12 (2013) 104, 1301.6949.

\bibitem{Siemens:2016jwj}
D. Siemens et~al.,
\newblock (2016), 1610.08978.

\end{thebibliography}

\end{document}